\documentclass[aps, prd, twocolumn, superscriptaddress, nofootinbib]{revtex4}
\usepackage{graphicx}
\usepackage{dcolumn}
\usepackage{amssymb,amsmath,amsthm,mathrsfs}
\usepackage{epstopdf}

\begin{document}

\newcommand{\Eq}[1]{Eq. \ref{eqn:#1}}
\newcommand{\Fig}[1]{Fig. \ref{fig:#1}}
\newcommand{\Sec}[1]{Sec. \ref{sec:#1}}

\newcommand{\PHI}{\phi}
\newcommand{\vect}[1]{\mathbf{#1}}
\newcommand{\Del}{\nabla}
\newcommand{\unit}[1]{\mathrm{#1}}
\newcommand{\x}{\vect{x}}
\newcommand{\ScS}{\scriptstyle}
\newcommand{\ScScS}{\scriptscriptstyle}
\newcommand{\xplus}[1]{\vect{x}\!\ScScS{+}\!\ScS\vect{#1}}
\newcommand{\xminus}[1]{\vect{x}\!\ScScS{-}\!\ScS\vect{#1}}
\newcommand{\diff}{\mathrm{d}}

\newcommand{\be}{\begin{equation}}
\newcommand{\ee}{\end{equation}}
\newcommand{\bea}{\begin{eqnarray}}
\newcommand{\eea}{\end{eqnarray}}
\newcommand{\vu}{{\mathbf u}}
\newcommand{\ve}{{\mathbf e}}
\newcommand{\vk}{{\mathbf k}}
\newcommand{\vx}{{\mathbf x}}
\newcommand{\vy}{{\mathbf y}}
\newcommand{\bx}{{\bf x}}
\newcommand{\bk}{{\bf k}}
\newcommand{\br}{{\bf r}}

\newcommand{\uden}{\underset{\widetilde{}}}
\newcommand{\den}{\overset{\widetilde{}}}
\newcommand{\denep}{\underset{\widetilde{}}{\epsilon}}

\newcommand{\nn}{\nonumber \\}
\newcommand{\dd}{\diff}
\newcommand{\fr}{\frac}
\newcommand{\del}{\partial}
\newcommand{\eps}{\epsilon}
\newcommand\CS{\mathcal{C}}

\def\be{\begin{equation}}
\def\ee{\end{equation}}
\def\ben{\begin{equation*}}
\def\een{\end{equation*}}
\def\bea{\begin{eqnarray}}
\def\eea{\end{eqnarray}}
\def\bal{\begin{align}}
\def\eal{\end{align}}

\def\TT{{\rm TT}}
\def\GW{{_{\rm GW}}}


\title{On the Anisotropy of the Gravitational Wave Background from Massless Preheating}

\newcommand{\addressImperial}{Theoretical Physics, Blackett Laboratory, Imperial College, London, SW7 2AZ, United Kingdom}

\author{Laura Bethke}
\affiliation{\addressImperial}

\author {Daniel G. Figueroa}
\affiliation{D\'epartement de Physique Th\'eorique and Center for Astroparticle Physics, Universit\'e de Gen\`eve, 24 quai Ernest Ansermet, CH--1211 Gen\`eve 4, Switzerland}

\author{Arttu Rajantie}
\affiliation{\addressImperial}


\date{\today}

\begin{abstract}
When a light scalar field is present during inflation, its value will vary on superhorizon scales, modulating the preheating process at the end of inflation. Consequently, the amplitude of the gravitational wave (GW) background produced during preheating will also be modulated. The observed energy density of this background will therefore appear anisotropic at different angles in the sky.
We provide a master formula for the angular power spectrum $C_l$ of the anisotropies in the GW background from preheating, valid for any scenario where the anisotropies are due to the superhorizon modulation of a light degree of freedom. 
Using lattice field theory simulations of massless preheating with $g^2/\lambda = 2$, we find a flat angular spectrum $l(l+1)C_l \approx 3\times10^{-4}$, which represents a 
strong anisotropy of $\sim1\%$ variations on large angular scales. For our choice of couplings, long wavelengths are amplified most strongly during parametric resonance, which is crucial for the development of the anisotropies. If future direct detection GW observatories are capable of detecting backgrounds of cosmological origin, they should be able to detect this effect. This could eventually become a powerful tool to discriminate among inflationary and preheating scenarios. 
\end{abstract}

 \pacs{To be done}

\maketitle


\section{Introduction}
The recent results from the Planck satellite~\cite{Planck} and other cosmological observations provide very strong support for the theory of cosmological inflation, according to which the Universe experienced an early stage of accelerated expansion. However, they offer little insight into the cause of this expansion; the observations are fully compatible with the simplest model with a single slowly rolling scalar field, but there are many other models whose predictions agree equally well with the data.

In order to learn more about the physics of the early universe, and of inflation in particular, one needs to consider new observables that could be used as probes of these primordial stages. One of the most promising candidates is gravitational radiation. On general grounds, it is expected that due to quantum metric fluctuations, an almost scale-invariant background of gravitational waves (GWs) was produced during inflation~\cite{inflationGW}. 
Furthermore, post-inflationary GWs are thought to have been significantly sourced by different non-equilibrium processes in the early universe. Specific backgrounds of GWs are predicted from bubble collisions in phase transitions~\cite{GWfromPhT,GWfromPhT2}, the creation~\cite{GWfromU1hybrid}, evolution~\cite{GWdefectsNetwork} and decay~\cite{GWstringsLoops} of cosmic defect networks, or from reheating~\cite{preheatingGW,preheatingGW2}, the transition from inflation to a radiation dominated universe. 

Depending on the physical parameters, each of these phenomena would give rise to a characteristic spectrum of GWs, which could be detected and measured by future experiments. The GW background from inflation, for instance, has either a flat or a tilted spectrum. Post-inflationary GW backgrounds, instead, have very specific spectral shapes, typically with some kind of feature such as a well defined peak, and characteristic slopes in the low- and high-frequency tails. In the case of GWs from reheating, the spectral shape contains relevant information about the corresponding inflationary model. While both the GW background from inflation and reheating can be used to extract the energy scale of inflation, the one from reheating also encodes information about the coupling(s) between the inflationary sector and other matter fields. This additional information is 'imprinted' in the details of the spectrum, i.e.~in its shape and height.

In a recent letter~\cite{LauraDaniArttuPRL}, we pointed out that in models with other light scalar fields beside the inflaton, the observed amplitude of the stochastic GW background from reheating would generally be anisotropic. The amplitude of the GWs would depend on the position where they were produced, and we would measure different amplitude
in different directions in the sky. 
The details of this anisotropy depend sensitively on the microscopic theory, and therefore potentially provide a very useful way to distinguish between different theories. The anisotropy pattern could be used as an additional observable on the inflationary sector, helping to break degeneracies in model parameters which would be indistinguishable if they were inferred from the spectral shape alone.

In typical inflationary models, preheating produces GWs at high frequencies, around $10^8-10^{11} {\rm Hz}$~\cite{preheatingGW,preheatingGW2}, which is far too high for current or planned experiments such as LIGO or NGO/eLISA. However, first efforts have been made towards building GW detectors at $\sim 10^8$ Hz frequencies~\cite{MHz}, although
the sensitivity of the existing prototypes is still far too low for detecting gravitational waves from preheating.
The relative amplitude of the anisotropy we find is of order $\sim 1\%$ on large angular scales, which means that if the GW background from preheating is ever detected, the sensitivity will only have to improve by two orders of magnitude in order to start probing the anisotropy.

In this paper we present full details of the calculations reported in Ref.~\cite{LauraDaniArttuPRL}. We also include some new data and carry out an improved analysis of the whole data set. The paper is organised as follows: In Section~\ref{massless} we define the massless preheating model and discuss its dynamics at a qualitative level. In Section~\ref{GWs}, we explain how GWs are produced during preheating. In Section~\ref{sepuni} we present general arguments why the observed GW amplitude can be anisotropic in the presence of light scalar fields. In Section~\ref{sims} we describe the numerical simulations we use to measure the GW anisotropy, and in Section~\ref{sec:Results} we present and discuss our results.
Finally, we conclude in Section~\ref{sec:conclusions}.

\section{Massless Preheating}\label{massless}

In many theories, reheating 
involves a highly out-of-equilibrium early stage driven by non-perturbative effects called preheating. The most common example is a period of parametric resonance between the inflaton field and other scalar fields~\cite{preheating,preheating2}. The resonance transfers a significant fraction of the inflaton's energy to the other fields, allowing the Universe to reheat much faster than through perturbative processes.

Massless preheating~\cite{GreeneKofmanLindeStarobinsky97} is a model where this phenomenon naturally occurs. It has the potential
\be \label{masslesspre} V(\phi,\chi) = \frac{\lambda}{4} \phi^4 + \frac{1}{2} g^2 \phi^2 \chi^2,\ee
where $\phi$ is the inflaton field, $\chi$ is another scalar field and $\lambda$ and $g$ are dimensionless coupling constants. Because the model has no dimensionful parameters, it is scale invariant and contains no fixed physical length scale. This makes it particularly convenient for numerical lattice field theory simulations.

The equations of motion for the two scalar fields in an expanding Friedmann-Robertson-Walker Universe, with scale factor $a(t)$, are
\bea
\ddot \phi + 3H\dot \phi - \frac{1}{a^2}\nabla^2\phi + (\lambda\phi^2 + g^2\chi^2)\phi &=& 0, \label{eomphi}\\
\ddot \chi + 3H\dot \chi - \frac{1}{a^2}\nabla^2\chi + g^2\phi^2\chi &=& 0, \label{eomchi}
\eea
where dots denote derivatives with respect to cosmic time, and $H=\dot{a}/a$ is the Hubble rate, determined by the Friedmann equation
\be H^2=\frac{4\pi}{3M_{\rm Pl}^2}\left[\dot{\phi}^2 + \dot{\chi}^2
+(\nabla\phi)^2 + (\nabla\chi)^2+2V(\phi,\chi)\right],
\label{Hubble}\ee
with $M_{\rm Pl} \simeq 1.22\times10^{19}$ GeV the Planck mass.
Note that any term on the the right-hand side of Eq.~(\ref{Hubble}) should be understood as spatially averaged. 

During inflation, $\phi$ has a large value and $\chi$ is small, and therefore the behaviour is the same as in $\lambda\phi^4/4$ chaotic inflation, a model that has now been clearly ruled out by observations~\cite{Planck}. Still, we choose to focus on it because of its computational convenience due to its scale invariance. 
Our aim is to compute the anisotropies arising in the GW background from preheating in this model. Later we will argue that these anisotropies are a common phenomenon arising naturally in other inflationary models, as long as certain conditions are met. Massless preheating is simply a good starting point for the analysis.

We are interested in the case in which the $\chi$ field is light during inflation, in the sense that its mass $m_\chi=g\phi$ is less than the inflationary Hubble rate $H$. Comoving modes of the $\chi$ field that leave the horizon while this condition is satisfied freeze out. This leads to a nearly scale-invariant Gaussian spectrum of fluctuations, with power spectrum~\cite{LiddleLyth}
\be\label{eq:ChiSpectrum}
{\cal P}_\chi \equiv {\partial\langle \chi^2 \rangle\over\partial\log k} 
\simeq {H^2\over4\pi^2},
\ee 
as for the inflaton field. At a time $N$ e-foldings before the end of inflation, the inflaton has the value $\phi=\sqrt{N/\pi}M_{\rm Pl}$~\cite{LiddleLyth}. Therefore, the field $\chi$ is light $N$ e-foldings before the end of inflation if
\be
\frac{m_\chi^2}{H^2}=\frac{3g^2\phi^2}{2\pi\lambda M_{\rm Pl}^2\phi^4}=\frac{3g^2}{2N\lambda}\lesssim 1.
\ee
In order for this to be the case for the largest observable scales, which left the horizon $N\sim 60$ e-foldings before the end of inflation, the couplings must satisfy $g^2/\lambda\lesssim 2N/3\sim 40$. 
If $H$ falls below $m_\chi$ before inflation ends, $\chi$ starts to oscillate with a decreasing amplitude, but if this underdamped period is short enough, the large-scale fluctuations of the $\chi$ field will still have a significant amplitude at the end of inflation and can lead to potentially observable effects. In practice we choose $g^2/\lambda=2$, which guarantees that $\chi$ is light apart from the very last moments of inflation.
For more details, see Appendix A.1 of Ref.~\cite{Chambers:2008gu}.

\begin{figure}[t]
\begin{center}
\includegraphics[width=8cm]{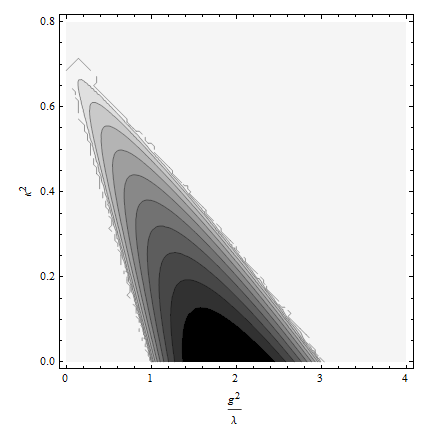}
\end{center}
\caption{Stability chart of the Lam\'e equation. Shaded regions correspond to unstable regions where fluctuations grow. The characteristic Floquet exponent $\mu(k,g^2/\lambda)$ is greater for darker regions, varying from $\mu \approx 0.2$ (darkest region), up to $\mu \approx 0.02$ (lightest region), in steps of $\Delta \mu = 0.02$.}
\label{fig:resonance}
\end{figure}

Ultimately, we will solve the full equations (\ref{eomphi})--(\ref{Hubble}) numerically, but it is instructive to consider the linearised approximation first. Ignoring both $\chi$ and the inhomogeneous modes of $\phi$, which are much smaller than the homogeneous $\phi$, Eq.~(\ref{eomphi}) reduces to
\be \ddot{\phi}+3H\dot{\phi} + \lambda \phi^3 = 0\,. \label{classinf}\ee
During inflation the friction term $3H\dot{\phi}$ dominates. Once inflation ends and the inflaton has rolled down the potential, the second derivative becomes important and $\phi$ oscillates around its minimum. Because it has no mass term, it has the equation of state of radiation, and therefore the Universe expands approximately as $a\propto t^{1/2}$~\cite{Turner1983}.
Using a conformal rescaling $\phi \rightarrow a \phi$ and rescaled conformal time $d\tau = (\sqrt{\lambda}\phi_i/a)dt$, the solution of equation (\ref{classinf}) can be written in the form of the Jacobi cosine function~\cite{GreeneKofmanLindeStarobinsky97}
\be\label{equ:linbg} \phi(\tau) = \phi_i \, \mbox{cn}(\tau, 1/\sqrt{2}) \, ,\ee
with $\phi_i$ the initial amplitude. 

The solution (\ref{equ:linbg}) has the same form as in non-expanding Minkowski space. This is due to the conformal invariance; the absence of any scale allows for a mapping into an equivalent problem in Minkowski space~\cite{GreeneKofmanLindeStarobinsky97}. The oscillating inflaton provides an oscillatory mass term for the field $\chi$, whose Fourier modes (again rescaled by $\chi \rightarrow a\chi$) obey the Lam\'e equation
\be \chi_k'' + \left[ \kappa^2+\frac{g^2}{\lambda} \mbox{cn}^2(\tau, 1/\sqrt{2})\right] \chi_k=0 \, ,\label{eq:chi_k}\ee
with $\kappa=k/(\sqrt{\lambda}\phi_i$) the rescaled wavenumber. Again, due to scale invariance, Eq.~(\ref{eq:chi_k}) has the same form as in Minkowski space. 

The inflaton self-interaction leads to the same mode equation for the fluctuations $\phi_k$, but with $g^2/\lambda$ replaced by the number 3. Thus, studying the fluctuations of both $\phi$ and $\chi$ amounts to analysing the same equation, Eq.~(\ref{eq:chi_k}). 

For any value of $g^2/\lambda$, the mode equation has unstable solutions for specific bands of momenta $\kappa$, for which the fluctuations grow exponentially as $\chi_k(\tau)=\exp\left[\mu(k,g^2/\lambda)\tau\right] f(\tau)$, where $f(\tau)$ is a periodic function and $\mu(k,g^2/\lambda)$ is a characteristic Floquet exponent quantifying the strength of the resonance. The resonance chart for the Lam\'e equation showing the stable and unstable regions was computed in Ref.~\cite{GreeneKofmanLindeStarobinsky97} and is partially reproduced in Fig.~\ref{fig:resonance}. For any value of $g^2/\lambda$ there are resonant comoving modes $\kappa$, but the strength of the resonance and which comoving wavelengths it affects depend sensitively on the ratio $g^2/\lambda$.

If $\chi$ is light, as we will assume, it acquires the spectrum given by Eq.~(\ref{eq:ChiSpectrum}). This 
implies that $\chi$ varies on superhorizon scales at the onset of preheating. Within each horizon volume, $\chi$ will have a distinct non-zero background value, which we will denote as $\chi_{\rm i}$.  The initial conditions at end of inflation are therefore set by a homogeneous amplitude $\chi_{\rm i}$, which is different in each horizon volume, 
superimposed with subhorizon vacuum fluctuations. 
The instability of the field during preheating will amplify the field modes that are inside the resonant bands for the given coupling ratio $g^2/\lambda$. If the resonance band includes $\kappa=0$, then the homogeneous mode will be amplified and can become significant by the time the dynamics become non-linear. 
This means that the non-linear evolution will take  place differently in each different Hubble patch, depending on the local value of $\chi_{\rm i}$, with important consequences for perturbations~\cite{Tanaka:2003cka,Suyama:2006rk,rajantie} and gravitational wave production~\cite{LauraDaniArttuPRL}.

\section{Gravitational Wave Production}\label{GWs}

Gravitational waves are produced during preheating as a result of a classical process, when fast moving inhomogeneities are generated in the scalar fields. Parametric resonance excites fluctuations only within a given characteristic range of momenta $k_*$, which depends on the couplings (see Fig.~\ref{fig:resonance}). This translates into field inhomogeneities in configuration space of size $L_* \sim 1/k_*$. 
The field distribution develops a rapidly evolving anisotropic stress, the transverse-traceless part of which acts as a very efficient source of GWs. The initial inhomogeneous configurations of size $L_*$ collide among themselves, breaking down into smaller inhomogeneities. This produces yet more GWs, but at smaller scales $k > k_*$. Eventually the fields relax, and the production of GWs ceases. In the end, a spectrum of GWs emerges with a certain width around $k_*$. Once produced, the GWs decouple from the matter fields, and propagate unimpeded until now. We should in principle be able to measure their spectrum today, and thus infer information about the process that generated them. 

Let us review how GWs are produced in more detail. During preheating, we consider tensor perturbations around a Friedmannian background,
\be
ds^2 = -dt^2 + a^2(t) (\delta_{ij}+h_{ij})dx^idx^j \,, \ee
with the transverse-traceless condition $\partial_i h_{ij}= h_{ii}=0$ satisfied, in order for $h_{ij}$ to be identified with GW degrees of freedom. Linearizing the Einstein equations, one finds an equation of motion for the perturbations, 
\be
\ddot h_{ij} + 3H\dot h_{ij} - \frac{1}{a^2}\nabla^2h_{ij} = \frac{16\pi}{M_{\rm Pl}^2a^2} \Pi_{ij}^\TT(\phi,\chi)\,, \label{tensoreq}\ee
where the source term for GWs, $\Pi_{ij}^\TT$, is the transverse-traceless (TT) part of the anisotropic stress tensor. In practice it depends only on the field gradients and can be written as~\cite{DaniThesis}
\be \Pi_{ij}^\TT = \left[ \partial_{i}\chi \,\partial_{j} \chi+ \partial_{i} \phi \, \partial_{j} \phi\right]^{\TT} \, ,\ee
where $[...]^{\TT}$ projects out the TT part of the expression inside the brackets. 

The TT-projection in configuration space is a non-local operation, so it is convenient to perform the projection in Fourier space. Defining a projector
\begin{eqnarray}
\Lambda_{ij,lm}(\hat k) = P_{il}P_{jm} - {1\over2}P_{ij}P_{lm}, \\
P_{ab} \equiv \delta_{ab} - k^{-2}k_ak_b\,,~~~~~~
\end{eqnarray}
the source of GWs can be written as
\begin{eqnarray}
\Pi_{ij}^\TT(\bk,t) = \Lambda_{ij,lm}(\hat k) \hspace*{5cm} \\ 
\times\int \hspace*{-1mm} d\bx \,e^{-i\bk\bx}\left\lbrace\partial_{l}\chi \,\partial_{m} \chi+ \partial_{l} \phi \, \partial_{m} \phi\right\rbrace\hspace*{-0.5mm}(\bx,t)\,.\nonumber
\end{eqnarray}
It is then guaranteed that $\Pi_{ii}^\TT(\bk,t) = k_i\Pi_{ij}^\TT(\bk,t) = 0$, $\forall\,\bk, t$.

For $g^2/\lambda \gtrsim 1$, the order of magnitude of the typical (dimensionless) momentum width  
$\Delta \kappa$ of the resonant modes is bounded
by~\cite{GreeneKofmanLindeStarobinsky97}
\begin{eqnarray}
\Delta \kappa 
\lesssim {1\over\sqrt{\pi}}\left({g^2\over \lambda}\right)^{1/4} \, .
\end{eqnarray}
This can be seen in Fig.~\ref{fig:resonance}. For $g^2/\lambda = 2$, the principal resonant band is $\Delta \kappa^2 \lesssim 0.3$, and thus the typical momenta are bounded as $\Delta\kappa \lesssim 0.55$, which is of the same order as $\pi^{-{1\over2}}2^{1\over4} \approx 0.67$. As also demonstrated by Fig.~\ref{fig:resonance}, for the particular choice of $g^2/\lambda = 2$, $\kappa = 0$ is the most resonant mode, since it has the greatest Floquet exponent. The resonance dies away as we approach $\kappa^2 \simeq 0.3$, i.e.~the Floquet index goes to zero. The spectrum of fluctuations, however, goes as $k^2|\chi_k|^2 \sim k^2e^{2\mu(k)}$, so its maximum amplitude will occur for some intermediate scale $\kappa_*$ between $0$ and $\Delta \kappa$, typically a fraction of $\Delta \kappa$. Therefore, we expect the spectrum of the fluctuations of the resonant field $\chi$ to be peaked at around a scale $\kappa_* \sim \mathcal{O}(0.1)$. The source of GWs, formed by products of fields, will therefore be peaked around $\kappa_*$ as well\footnote{Due to the gradients in the anisotropic stress expression, the GW source term should actually be peaked at a slightly different scale than $\kappa_*$. However, this is just a small shift, so the actual peak scale is expected to be of the same order as $\kappa_*$.}. The tensor perturbations will then inherit this scale, and correspondingly the spectrum of GWs should also be peaked around $\kappa_*$. 

The total GW energy density within a volume $V = L^3$, normalised to the critical energy density $\rho_c$ is defined as
\be \Omega_\GW(t) = \frac{1}{\rho_c} \int \left(\frac{d\rho_\GW}{d\log k}\right) d\log k\,, \ee
with the GW power spectrum per logarithmic momentum interval given by~\cite{DaniJuanArttu}
\be \frac{d\rho_\GW}{d\log k}(k,t) \equiv \frac{k^3M_{\rm Pl}^2}{(4\pi L)^3} \int {d\Omega_k\over 4\pi} \dot{h}_{ij}(t,k,\hat\vk)\dot{h}^{\ast}_{ij}(t,k,\hat\vk)  \label{GWspect}\ee
More explicitly, the tensor modes can be written in terms of an appropriate Green function $\mathcal{G}(k,t-t')$ as
\begin{eqnarray}\label{eq:h_ijGreenFunc}
\dot h_{ij}(\bk,t) = {16\pi k\over M_{\rm Pl}^2} \int^t {dt'}\,\mathcal{G}(k(t-t'))\,\Pi_{ij}^{\TT}(\bk,t')\,,
\end{eqnarray} 
while the source term in Fourier space is
\begin{eqnarray}\label{eq:PI_ijExplicit}
\Pi_{ij}^{\TT}(\bk,t) = \Lambda_{ij,lm}(\hat\bk)\int \hspace*{-1.5mm}d{\bf q}\,\,q_l\,q_m\, \chi_{q}(t)\,\chi^*_{|\bk-{\bf q}|}(t)\,,
\end{eqnarray}
plus an analogous expression for the fluctuations of $\phi$. Looking at the last three equations we can now better understand the explanations above: $\Pi_{ij}^\TT(\bk,t)$ will be peaked at $k_*$ since it depends directly on the $\chi$ fluctuations through Eq.~(\ref{eq:PI_ijExplicit})
. From Eq.~(\ref{eq:h_ijGreenFunc}) we see that $\dot{h}_{ij}(k,\hat\bk,t)$ will inherit the peak scale from $\Pi_{ij}^\TT(\bk,t)$, and consequently so will $\frac{d\rho_\GW}{d\log k}(k,t)$ via Eq.~(\ref{GWspect}). 

\begin{figure}[t]
\begin{center}
\includegraphics[width=6.5cm]{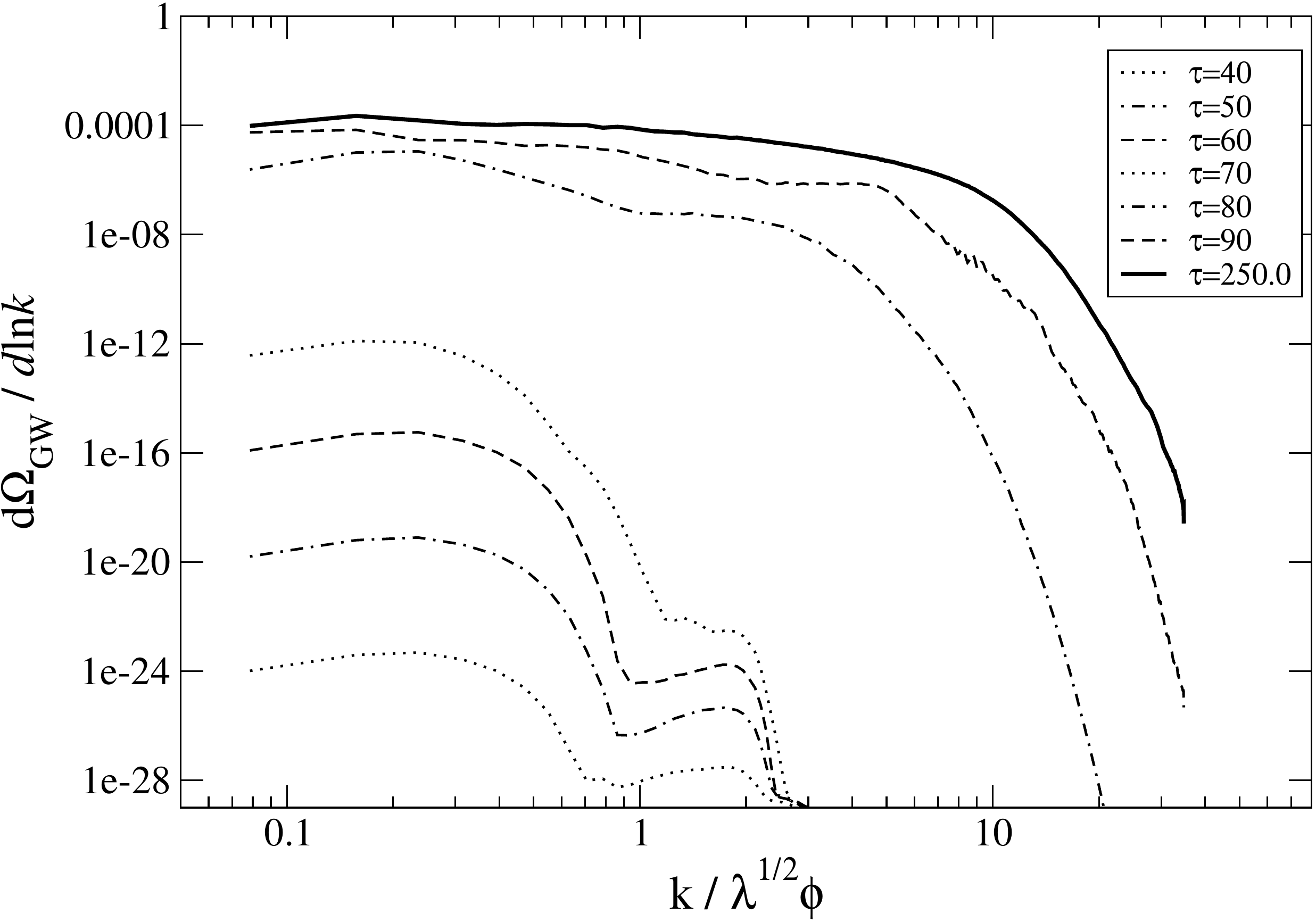}
\end{center}
\caption{Typical GW spectra from massless preheating with $g^2/\lambda = 2$, shown at different time steps as the amplitude grows. The highest curve (continuous line) corresponds to the final time step of our simulation $\tau=250$, when the amplitude saturates. The peak of the spectrum is at $\kappa_* \sim \mathcal{O}(0.1)$. The production of GWs increases significantly between $\tau=70$ and $\tau=80$, when the system becomes non-linear and there is a transfer of power into smaller scales (higher momenta).}
\label{GWspectll}
\end{figure}

We thus expect the energy spectrum of the GW background from massless preheating to be peaked around a scale slightly smaller than $\Delta \kappa \sim \pi^{-1/2}(g^2/\lambda)^{1/4}$. In Fig.~\ref{GWspectll} we show the evolution in time of such spectrum obtained for $g^2/\lambda = 2$. At early times, the peaks are located around $\kappa_* \approx 0.25$, which is about a factor three smaller than $2^{1/4}/\sqrt{\pi} \approx 0.67$. The amplitude of the GWs grows due to the initial parametric resonance until non-linearilities appear, increasing the amplitude further and transferring power into the higher momentum modes.

\section{Anisotropic Gravitational Wave Background}
\label{sepuni}

As mentioned in Section~\ref{massless}, the lightness of the scalar $\chi$ implies that, at the end of inflation and the time of preheating, $\chi$ will vary on superhorizon scales. In each horizon volume, $\chi$ will have a distinct non-zero background 
value $\chi_{\rm i}$. Any quantity $f$ that depends on the value of $\chi_{\rm i}$, $f(\chi_{\rm i})$, will consequently vary between different preheating horizon volumes. 

In each Hubble volume the unstable resonant modes will grow very similarly as long as the evolution is linear. If the homogeneous mode $\kappa=0$ is inside a resonance band, it will also grow exponentially and can be significant by the time the system becomes non-linear. In that case, the whole non-linear evolution can depend sensitively on the initial value $\chi_{\rm i}$ of the homogeneous mode. Therefore it will also have a strong effect on the production of GWs. As the effect arises from non-linear dynamics, it cannot be computed analytically.

In our current observable universe, the GW background from preheating detected on Earth today would have originated from a comoving spherical shell of radius $R\sim 1/H_0$, where $H_0$ the Hubble rate today. 
This shell clearly contains a very large number of preheating Hubble volumes. Any direction ${\hat n}$ on the sky will be pointing to one of these primeval volumes centered at ${\bf r}=R{\hat n}$. 
As $\chi$ varies on cosmological scales, and since we expect that $\Omega_{\GW}$ is a function of $\chi_{\rm i}$, the GW amplitude will then depend on the direction ${\hat n}$, i.e.~$\Omega_{\GW}({\hat n})= \Omega_{\GW}(\chi_{\rm i}(R{\hat n}))$. In other words, the measured amplitude of the gravitational wave background from preheating will be anisotropic on the sky.

The question then is, how large is the difference in amplitude in the GW background when observed at different angular directions in the sky? To answer this, we will study preheating numerically, considering a collection of preheating Hubble volumes (which are in causal contact in our current observable universe, but evolved independently during preheating), with different initial values $\chi_{\rm i}$. This separate universe approximation was employed by \cite{rajantie,Chambers:2008gu} in the context of massless preheating, to study the effect of $\chi_{\rm i}$ on the curvature perturbations. Later, more accurate calculations \cite{Bond} showed that certain initial values $\chi_{\rm i}$ lead to spikes in the curvature perturbation from inflation, which would result in cold spots in the CMB. Here we will use the same approach to calculate the dependence of $\Omega_{\GW}$  on $\chi_{\rm i}$. 

In order to do that, we first need to determine what range of $\chi_{\rm i}$ values we can expect the GW background from preheating to have originated from. Since $\chi_{\rm i}$ is a Gaussian random field with a scale-invariant spectrum (\ref{eq:ChiSpectrum}), it will have a non-zero average value in any given volume, even in the comoving volume that corresponds to the currently observable Universe. The range of comoving wavelengths that are amplified by inflation extends from the Hubble length at the end of inflation, $k\sim H_*$, which is well inside the horizon today, to the Hubble length at the start of inflation, which probably corresponds to a superhorizon scale much larger than our current horizon.

From the observational point of view, the wavelengths that are currently inside the horizon, $k\gtrsim a_0H_0$ appear as inhomogeneous fluctuations, or anisotropies on the sky.
The variance $\sigma_\chi^2$ of these fluctuations can be computed from the power spectrum (\ref{eq:ChiSpectrum}),
\be
\label{chivar}
\sigma_\chi^2=
\int_{H_*}^{a_0H_0} \frac{dk}{k}\,{\cal P}_\chi = \frac{H_*^2}{4\pi^2}\,N_*\,,
\ee
where $N_*\sim 60$ is the number of e-folds after the largest observable scales left the inflationary Hubble radius, and we have fixed $a_* = 1$ at the end of inflation.

If inflation lasted longer than $N_*\sim 60$ e-folds, even larger scales had been amplified and $\chi$ will have varied on scales that are superhorizon now. The actual mean value $\overline\chi_{\rm i}$ across the Universe would be a particular realization drawn from a Gaussian distribution with variance
\bea
\label{chimean}
\langle{\overline\chi}_{\rm i}^2\rangle &=& \int_{a_0H_0}^{(aH)_{\rm start}} \frac{dk}{k}\,{\cal P}_\chi \nonumber\\
&=& \int_{H_*}^{(aH)_{\rm start}}\frac{dk}{k}\,{\cal P}_\chi -\int_{H_*}^{a_0H_0}\frac{dk}{k}\,{\cal P}_\chi\nonumber\\
&=& \frac{H_*^2}{4\pi^2}\,(N_{\rm tot}-N_*)\,,
\eea
where $N_{\rm tot}=\ln(1/a_{\rm start})$ is the total number of e-foldings of inflation.  A typical average field value across a volume as large as our observable Universe is then
\begin{eqnarray}
\overline{\chi}_{\rm i} \sim \frac{H_*}{2\pi}\sqrt{(N_{\rm tot}-N_*)}\,.
\label{chitypical}
\end{eqnarray}
Since the value of $N_{\rm tot}$ is unknown, we will consider the actual realization of $\overline{\chi}_{\rm i}$ within our observable patch as a free parameter, simply restricted to $\overline{\chi}_{\rm i} > H_*/2\pi$. 

We will study how $\Omega_\GW$ depends on the different values of $\chi_{\rm i}$ as drawn from a Gaussian distribution with variance $\sigma_\chi^2$ given by Eq.~(\ref{chivar}), and centered around a mean value $\overline \chi_{\rm i}$ of our choice [but of the order of Eq.~(\ref{chitypical})].

\section{Simulations}\label{sims}

We have used comoving 3d lattice simulations to study the production of GWs during preheating numerically. Our code was based on the publicly available MPI C/C++ ClusterEasy package~\cite{CLUSTEREASY}, which uses a second-order leap-frog integrator with periodic boundary conditions. The accuracy of this method was sufficient for our purposes (as opposed to the case where the contribution of $\chi$ to curvature perturbations was analysed as in \cite{rajantie,Bond}, where a more accurate integrator was needed). The code solves discretized versions of the field equations for the scalars, Eqs.~(\ref{eomphi}), (\ref{eomchi}), and the Friedmann equation Eq.~(\ref{Hubble}). To compute the GW spectrum we need to solve the equation for the tensor perturbations, Eq.~(\ref{tensoreq}), and calculate Eq.~(\ref{GWspect}). However, performing the TT projection on the GW source and Fourier transforming to momentum space and back at all time steps of the simulation would be computationally very costly. To avoid having to do this, we follow the method introduced by~\cite{Figueroa1}, which amounts to the following procedure. In Fourier space, Eq.~(\ref{tensoreq}) has the formal solution given by Eq.~(\ref{eq:h_ijGreenFunc}), as long as the initial conditions are $h_{ij}(k,t_i) = \dot h_{ij}(k,t_i) = 0$. The solution can be re-written as 
\begin{eqnarray}\label{eq:PhysicalSolution}
\dot h_{ij}(k,t) = \Lambda_{ij,lm}(\hat k)\dot u_{lm}(k,t)
\end{eqnarray}
where
\begin{eqnarray}\label{eq:FakeSolution}
\dot u_{lm}(k,t) \equiv {16\pi k\over M_{\rm Pl}^2} \int^t {dt'}\,\mathcal{G}(k(t-t'))\,\Pi_{lm}^{\rm eff}(\bk,t')\,,
\end{eqnarray}
with $\Pi_{ij}^{\rm eff}(\bk,t)$ the Fourier transform of
\begin{eqnarray}\label{eq:EffectiveSource}
\Pi_{ij}^{\rm eff}(\bx,t) \equiv \left(\partial_i\chi\partial_j\chi + \partial_i\phi\partial_j\phi\right)(\bx,t) \, ,
\end{eqnarray}
the unprojected source term.
The idea is to solve a discretized version of the equation of motion
\be
\ddot u_{ij} + 3H\dot u_{ij} - \frac{1}{a^2}\nabla^2u_{ij} = \frac{16\pi}{M_{\rm Pl}^2a^2} \Pi_{ij}^{\rm eff}(\phi,\chi)\,
\label{eq:tensorFakeeq}
\ee
in configuration space. Only at the times for which we actually want to obtain the GW spectrum, we then Fourier transform $\dot u_{ij}(\bx,t)$ to $\dot u_{ij}(\bk,t)$, and recover the real GW degrees of freedom $\dot h_{ij}(\bk,t)$ by means of the projection in Eq.~(\ref{eq:PhysicalSolution}). From there, we can simply compute the spectral amplitude for each mode by using Eq.~(\ref{GWspect}). 

On a lattice, there is some ambiguity regarding the discretization of the projection operator $\Lambda_{ij,lm}$ in Eq.~(\ref{eq:PhysicalSolution}). The different projections, based on the different discretized schemes for lattice derivatives, were analysed in detail in \cite{DaniJuanArttu}. In our simulations we used a discretized version of Eq.~(\ref{GWspect}) given by Eq.(4.5) in~\cite{DaniJuanArttu}. We chose to use a simple real projector for which the transversality condition is attained by neutral lattice derivatives defined as $\Delta_i f[n] \equiv {1\over 2dx}[f(n + \hat i dx) - f(n - \hat i dx)]$, where $n$ is a lattice site and $dx$ the lattice spacing. We checked that the choice of projector did not influence our results.

The only parameters in the model we need to specify are the two coupling constants, and in particular their ratio $g^2/\lambda$. The CMB normalization fixes the inflaton self/coupling to $\lambda=9 \times 10^{-14}$~\cite{LindeBook}. We take $g^2/\lambda=2$ to ensure that the field $\chi$ is light during inflation, and that its longest wavelength modes get amplified strongly (Section \ref{massless}). We also ran simulations for $g^2/\lambda = 1$ and $6$, which also guarantee the lightness of $\chi$ during inflation, but do not lead to the amplification of homogeneous modes. In such cases we did not observe any anisotropies, which shows that the long wavelength mode dynamics are very important for the production of GWs.

At the start of every simulation, we chose the following initial conditions: the scale factor was set to $a_i = 1$, and the inital amplitude of the homogeneous inflaton to $\phi_i=0.342M_{\rm Pl}$, corresponding to the value for which $\dot\phi_i = -H_*\phi_i$ in the slow roll regime. Fluctuations in the fields $\phi$ and $\chi$ were set mimicking quantum vacuum fluctuations: considering each mode ($\phi_k$, $\chi_k$) given by a complex number $|f_k|e^{+i\varphi_k}$, the phases $\varphi_k$ were taken as random numbers uniformly distributed between $[0,2\pi)$, while the moduli amplitudes were set according to a Rayleigh distribution~\cite{khlebnikovtkachev}, with variance 
\begin{eqnarray}
\langle |f_k|^2 \rangle = {1\over2a^2 \omega_k}\,,~~~~~ \omega_k \equiv \sqrt{k^2 + m_f^2},
\end{eqnarray}
and effective masses $m^2_\phi \equiv 3\lambda\phi_i^2 + g^2\chi_{\rm i}^2$ and $m^2_\chi \equiv  g^2\phi_i^2$. The initial background value $\chi_{\rm i}$ was chosen as described in section \ref{GWs}. We considered a Gaussian random distribution with variance given by Eq.~(\ref{chivar}) and mean value by Eq.~(\ref{chitypical}). From Eq.~(\ref{Hubble}) the Hubble rate at the end of inflation (when the first derivative term can be neglected) is given by 
\begin{eqnarray}
H_{\ast}^2\approx {8\pi\lambda\phi_i^4\over12 M_{\rm Pl}^2} \simeq 2.6\times 10^{-15}M_{\rm Pl}^2.
\end{eqnarray}
Using $N_*\sim 60$ and Eq.~(\ref{chivar}), the variance of $\chi_{\rm i}$ across the observable universe is then $\sigma_\chi^2 \simeq 4\times 10^{-15}M_{\rm Pl}^2$. Taking $(N_{\rm tot}-N_*) \sim 100$, the mean value of $\chi_{\rm i}$ is of order $\overline{\chi}_{\rm i}\sim 10^{-7}M_{\rm Pl}$. In our simulations we made the specific choice $\overline{\chi}_{\rm i} = 3.42\times10^{-7}M_{\rm Pl}$, and later extrapolated our results to neighboring values by the Monte Carlo reweighting method, as we explain in Section~\ref{subsec:Anisotropies2}.

An important program variable to specify is the lattice size $L$, which ultimately fixes the infrared momentum cutoff $k_{\rm IR}=2\pi/L$, i.e.~the minimum momentum captured by a simulation. For the case $g^2/\lambda=2$, the long wavelength modes get amplified most strongly, so we need to choose quite a large lattice volume to ensure that we capture the peak of the GW spectrum. At low $k$, the $k^3$ term in Eq.~(\ref{GWspect}) dominates, showing that there is no amplification of modes on scales that are not causally connected. In other words, despite $k = 0$ being the maximally amplified resonant mode, the maximum amplitude of the GW spectrum is at some scale $k_* \neq 0$, as discussed in Section~\ref{GWs}. We therefore need to ensure that our choice of lattice volume is such that the IR cut-off satisfies $k_{\rm IR} < k_*$.

In the natural program variables, the comoving lattice size $L$ is given in units of $(\sqrt{\lambda}\phi_i)^{-1}$, so that we can define a dimensionless variable $\widetilde L \equiv \sqrt{\lambda}\phi_iL$. For the choice of couplings $g^2/\lambda = 2$, the peak of the resonant modes, and thus of the GWs, is located around $\kappa_* \sim 0.1$ (see Fig.~\ref{fig:GWspectra}). This forced us to consider $\widetilde L \sim \mathcal{O}(100)$ lattice sizes, in order to guarantee that the largest excited wavelength modes were captured. We found that any value between $\widetilde L = 80$ and $\widetilde L = 160$ turned out to be reasonable choice, despite the fact that they initially exceed the comoving Hubble radius. Ideally, the simulation volume should be smaller than the comoving Hubble volume at all times, to make sure we only describe causally connected regions. However, the assumption of a Friedmannian background on scales that are larger than the Hubble horizon is justified as long as no variations in the scale factor, i.e. curvature perturbations, are being considered. Furthermore, the largest wavelength modes in the chosen lattice volume(s) turn into subhorizon during the stage of maximal GW production, so in this sense our choice of dynamical range is justified. 

\begin{figure}[t]
\begin{center}
\includegraphics[width=8.0cm]{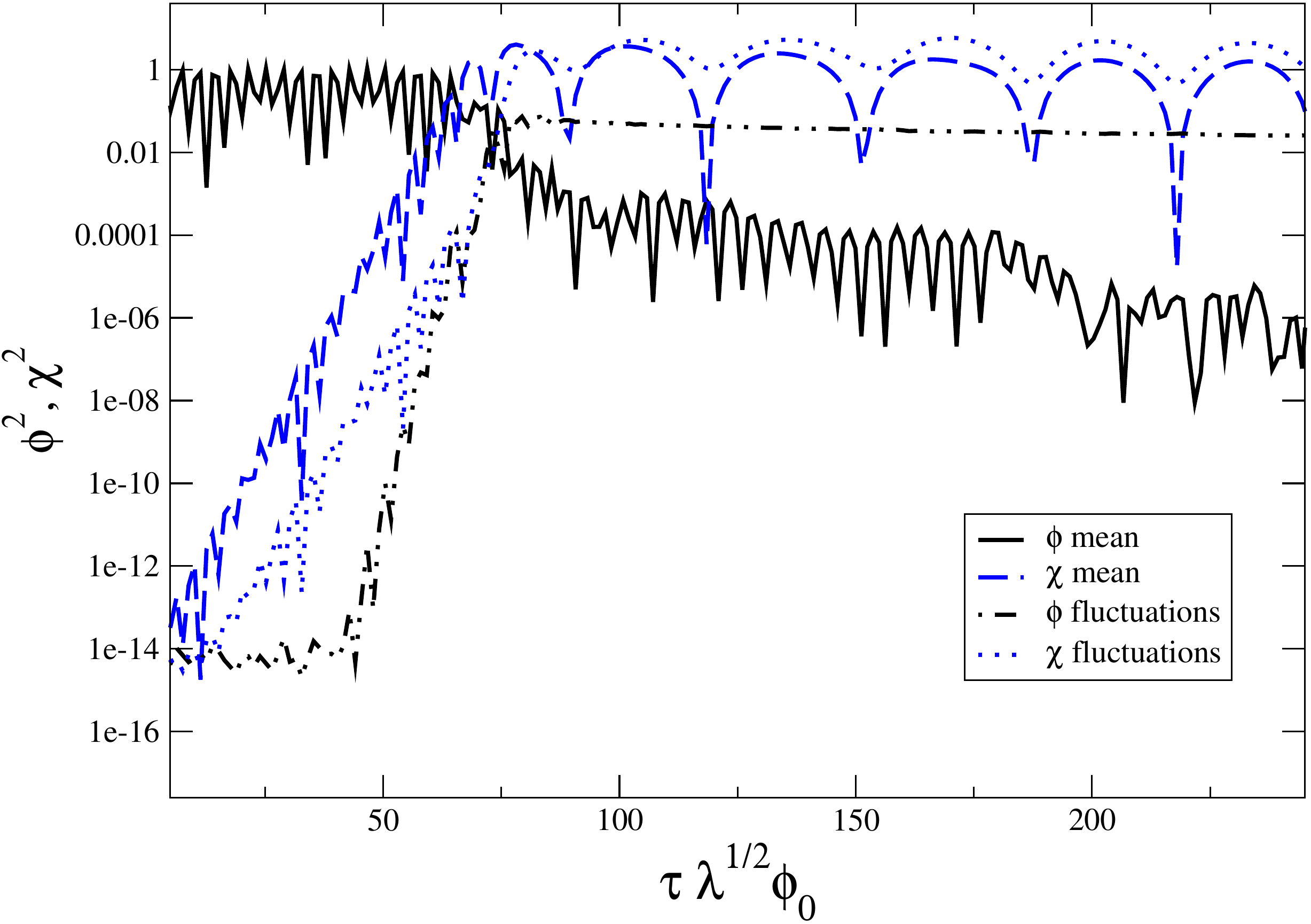}
\end{center}
\caption{Evolution of the of the mean field amplitudes squared and of their variances. The exponential growth of fluctuations due to parametric resonance can be clearly appreciated. }
\label{meansvar}
\end{figure}

To also ensure a good coverage of the short wavelength ultraviolet modes, it was necessary to use lattices with a large number $N$ of points per dimension, at least $N = 512$. The GW spectrum plotted in Fig.~\ref{GWspectll} shows the large dynamical range captured by our simulations, where the momenta span more than two orders of magnitude.

Our simulations confirm the usual behaviour of parametric resonance, see Fig.~\ref{meansvar}. Note that here and in all later discussions fields have been rescaled by the scale factor $a$. Initially the amplitude of $\phi$ is much larger than that of $\chi$, but the oscillations of the former induce a resonant growth of the $\chi$ fluctuations. This is shown very clearly by the variance term $\langle \chi^2 \rangle$, which grows exponentially fast from from $\tau = 0$ to $\tau = 70$, where $\tau$ is the rescaled conformal time $\tau = (\sqrt{\lambda}\phi_i/a)t$. The variance in $\phi$ grows as well due to its self-interactions and coupling to $\chi$, but its growth only starts at around $\tau = 40$, once $\langle \chi^2 \rangle$ has already been amplified by around six orders of magnitude. The energy transferred from $\phi$ to $\chi$ is significant, so the (mean) amplitude of $\chi$ eventually reaches that of $\phi$, at about $\tau=70$, and the system becomes non-linear. For further details on the linear dynamics of the system, we refer the reader to Ref.~\cite{GreeneKofmanLindeStarobinsky97}.

\begin{figure}[t]
\begin{center}
\includegraphics[width=8cm]{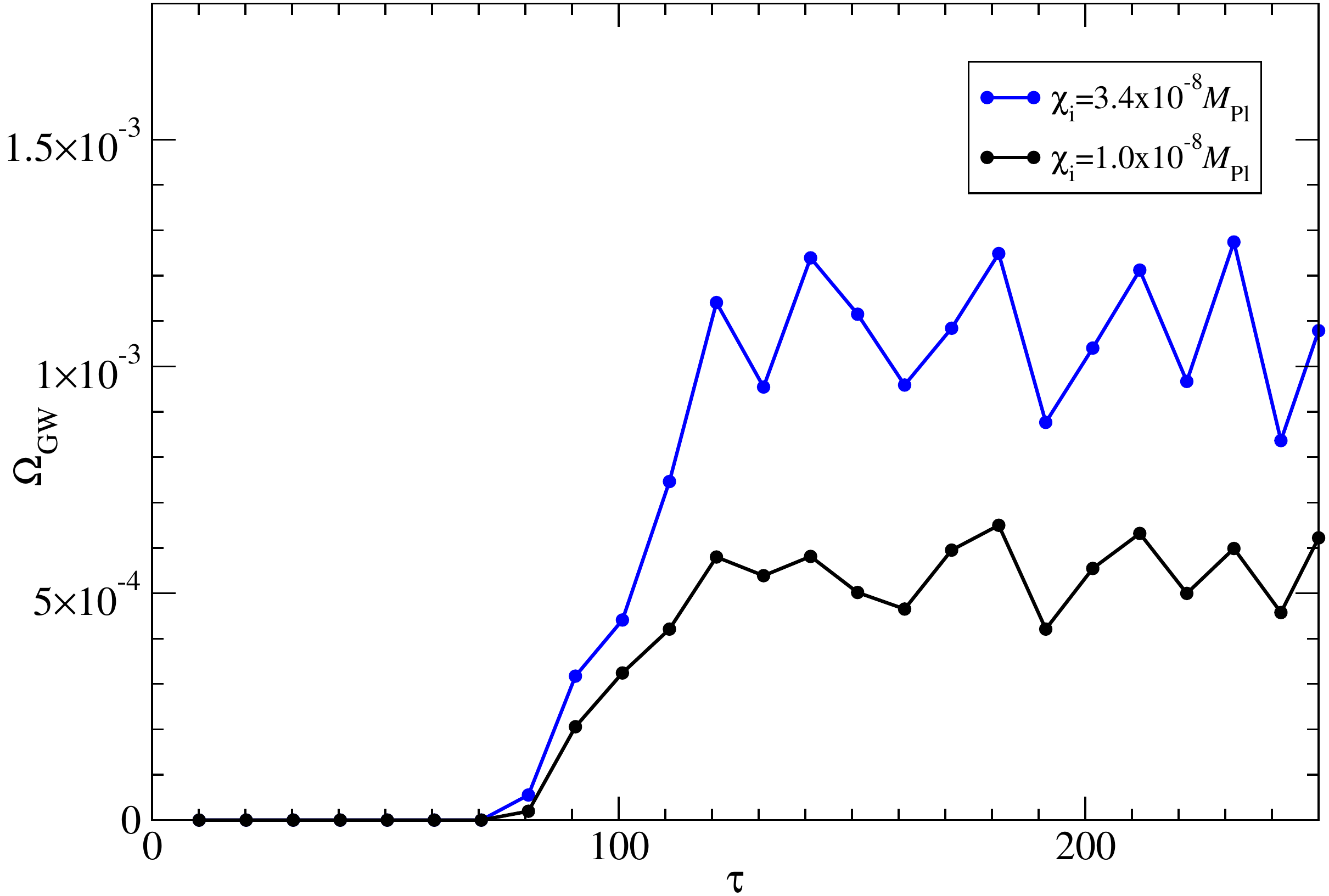}
\caption{
The total energy density of gravitational waves as a function of rescaled conformal time $\tau$ for two different initial field values $\chi_{\rm i}$.
\label{fig:GWevolution}
}
\end{center}
\end{figure}

The production of GWs starts during the initial stage of exponential growth of the $\chi$ fluctuations, between $\tau = 0$ and $\tau = 70$. During the subsequent stage of non-linear evolution, from $\tau = 70$ until $\tau = 100$, the field gradients become much larger, and consequently GWs are being produced with larger intensity. This can be seen in Fig.~\ref{GWspectll}. The GW production finally reaches an end when the system enters into a turbulent regime, on its way towards equilibration. 
The time evolution of the total GW energy density, obtained by summing over all lattice momenta, is shown in Fig.~\ref{fig:GWevolution}. Because of turbulent oscillations of the fields, it oscillates around a constant mean value at the end of simulation. Therefore we need to average over a few oscillations to obtain a stable final value 
for the gravitational wave energy density $\Omega_\GW$.

The final amplitude of the GW spectrum, obtained from an average over several time oscillations as well as over five random realisations of the inhomogeneous modes, is shown in Fig.~\ref{fig:GWspectra}. We will comment on the significance of the two different initial values $\chi_{\rm i}$ in section \ref{subsec:GW_vs_InitialChi}.  Note that the error bars only contain the meaningful statistical variation due to different seeds (but we want to point out that the fluctuations in time were of the same order of magnitude). The dashed lines corresponds to our fiducial choice of lattice size and number of points per dimension $(\widetilde L, N) = (80,512)$, whereas the solid lines correspond to $(\widetilde L, N) = (160,1024)$, ensuring the same ultraviolet (UV) coverage. For $\widetilde L=160$, one can clearly see a drop in the infrared (IR), which shows that very long wavelength modes are not excited, as expected from causality. The runs with $(\widetilde L, N) = (160,1024)$ were computationally too expensive for our purposes, which require performing several hundreds of simulations as we shall explain in Section~\ref{subsec:Anisotropies}. However, we chose to run a few simulations with such a large lattice volume to show that, in practice, the total integrated GW amplitude from the two cases 
agrees to better than $1\%$. This demonstrates that the fiducial case $(\widetilde L, N) = (80,512)$, which we use systematically in Section~\ref{subsec:Anisotropies}, is not dominated by lattice artifacts, and therefore meets our needs. For lattices with $N = 256$, independently of the volume $\widetilde L$, it was not possible to capture both the IR and UV behaviour sufficiently well at the same time. Runs with $(\widetilde L, N) = (> 80,512)$ improved the IR coverage but would require to upgrade to $N = 1024$ to keep a sufficiently good UV coverage, which, as mentioned before, was too costly computationally. The choice $(\widetilde L, N) = (80,512)$ therefore turned out to be the optimal one for our goals, representing a good compromise between a sufficiently large dynamical range, and low enough memory usage and shorter duration of the runs.

\begin{figure}[t]
\begin{center}
\includegraphics[width=8.0cm]{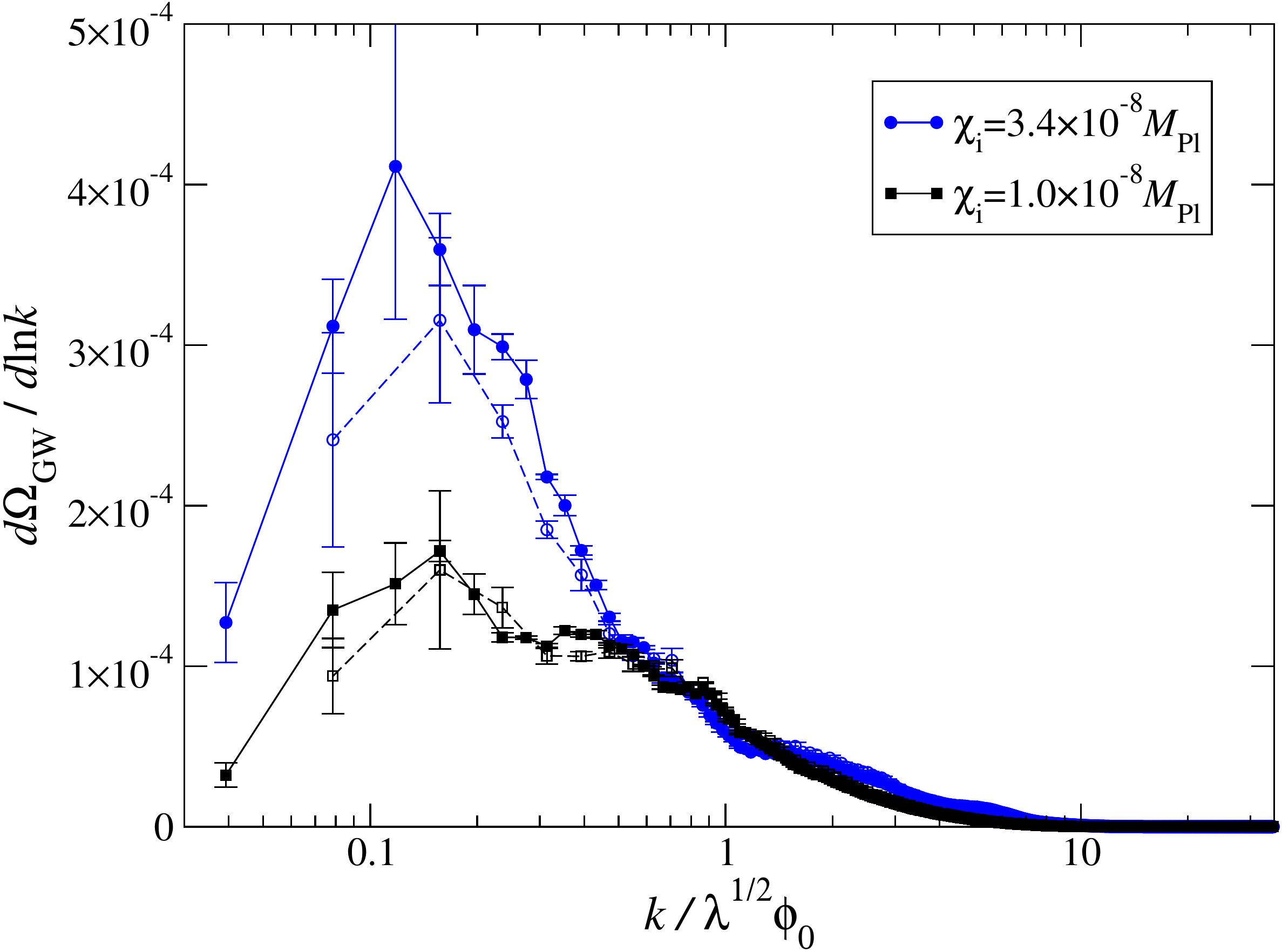}
\end{center}
\caption{Final spectrum of GW for 
$\chi_{\rm i} = 3.4\times 10^{-8}M_{\rm Pl}$ 
(upper, blue curves)
and 
$\chi_{\rm i}=1.0\times 10^{-8}M_{\rm Pl}$ 
(lower, black curves), 
averaged over time oscillations and five random realizations of inhomogeneous fluctuations. The solid curves are for $\tilde{L}=160$, $N=1024$, and the dashed curves for $\tilde{L}=80$, $N=512$. The area underneath corresponds to the total fractional GW energy density within a preheating Hubble domain.}
\label{fig:GWspectra}
\end{figure}

\section{Results}
\label{sec:Results}

In this section we will describe and quantify the anisotropy pattern arising in the GWs from massless preheating in detail. We will first consider the impact on the dynamics of $\chi$ due to different $\chi_{\rm i}$ values as an initial condition. Different behaviour in the $\chi$ dynamics depending on $\chi_{\rm i}$ will then give rise to a different amplitude in the produced GWs, thus creating an inhomogeneous distribution of the GW energy density over cosmological scales. We will show how to treat the produced inhomogeneities statistically, which today will be perceived as an anisotropic variation in the amplitude of the GW spectrum, depending on the direction of observation. We will quantify the amplitude and shape of the angular power spectrum of the relative fluctuations in the total energy density of GWs, and analyse its dependence on the actual mean value $\overline \chi_{\rm i}$ realised in our observable Universe. We will end the section by providing further insight into the physical origin of the anisotropy pattern arising in the GWs from massless preheating.  

\subsection{The impact of $\chi_{\rm i}$}
\label{subsec:GW_vs_InitialChi}

Let us begin by observing how different values of $\chi_{\rm i}$ give rise to a different GW background from massless preheating. Fig.~\ref{fig:GWspectra} shows the GW spectrum obtained from simulations initialised with two different values, $\chi_{\rm i} = 3.4\times 10^{-8}M_{\rm Pl}$ (upper, blue curve) and $\chi_{\rm i}=1.0\times 10^{-8}M_{\rm Pl}$ (lower, black curve), chosen for the purpose of illustration. 
As Fig.~\ref{fig:GWspectra} shows a log-linear plot, the area underneath the curves corresponds to the total fractional GW energy density within a preheating volume. While the peaks of the spectra are located at the same scale $\kappa_* \simeq 0.1-0.2$ (which is, as mentioned previously, just determined by the scale at which fluctuations get amplified), the amplitudes are not comparable. The two initial $\chi_{\rm i}$ values therefore lead to a large difference in the GW energy density, with the peaks of the spectra differing in amplitude by a factor of four. This $\mathcal{O}(1)$ effect is much larger than what could be naively expected.

\begin{figure}[t]
\begin{center}
\includegraphics[width=10.0cm]{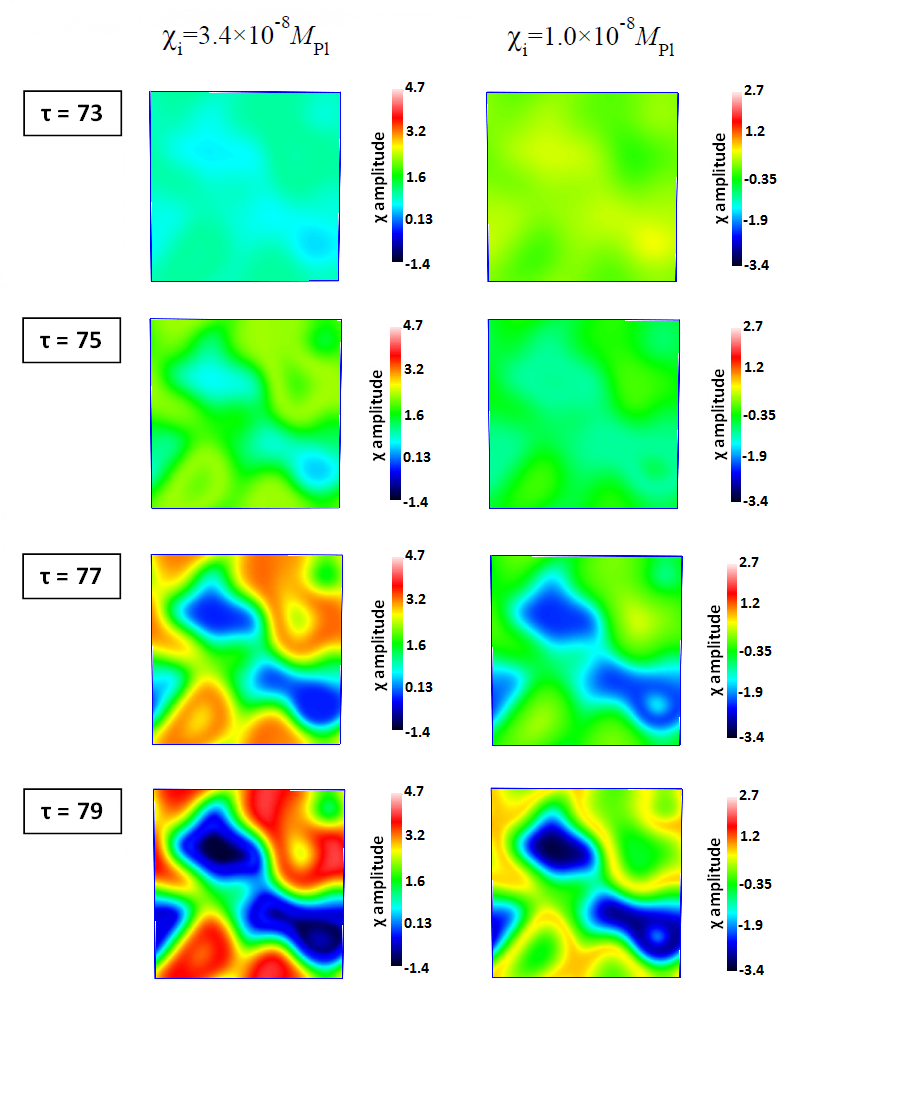}
\end{center}
\caption{2d snapshots of the 3d distribution of $\chi$ at different times of the evolution during preheating, from $\tau = 73$ to $\tau = 79$, the time when the GWs are being sourced most actively. The left panels correspond to the case $\chi_{\rm i} = 3.42\times 10^{-8} M_{\rm Pl}$, and the right panels to $\chi_{\rm i} = 1.0\times 10^{-8} M_{\rm Pl}$. The color coding is fixed during the evolution, though different between the two cases. However, the range of $\chi$ values covered by the axis is the same in both cases, such that different colours describe the same magnitude of difference in both cases. The correlation between the dynamics of the sources and the amplitude of the GWs is clearly demonstrated by this sequence of snapshots: the gradients for $\chi_{\rm i} = 3.42\times 10^{-8} M_{\rm Pl}$ are larger than for $\chi_{\rm i} = 1.0\times 10^{-8} M_{\rm Pl}$, in correspondence with the higher amplitude of GWs, as shown in Fig.~\ref{fig:GWspectra}. }
\label{fig:2d_snapshots}
\end{figure}

Since the two cases in Fig.~\ref{fig:GWspectra} start with different random initial conditions for the UV modes, one might be tempted to think such a difference in amplitude could be just a statistical fluctuation. However, the error bars in each plot of the figure, which correspond precisely to the variation of different initial random seeds for the initial quantum fluctuations in $\chi$, demonstrate that this is not the case: the magnitude of the statistical fluctuation is much smaller than the variation in the GW amplitude due to considering two different initial $\chi_{\rm i}$ values. The final discrepancy in amplitude of the GW spectra must therefore arise because of the different behaviour of the fields sourcing the GWs, ultimately due to the different choice of initial amplitude $\chi_{\rm i}$.

Because the GWs are sourced by field gradients, the homogeneous component has no effect until the evolution becomes nonlinear. However, if the homogeneous mode lies inside a resonance band, as in our case, it grows exponentially and contributes significantly to the nonlinear dynamics. 
Different values of $\chi_{\rm i}$ will therefore create a different outcome in the spatial distribution of $\chi$. 

As explained in Section~\ref{GWs}, the fact that parametric resonance only occurs at finite bands in momentum space produces a very inhomogeneous distribution in configuration space. 
In Fig.~\ref{fig:2d_snapshots}, we show a time sequence of 2d snapshots of the 3d spatial distribution of the field $\chi$. We compare the same values $\chi_{\rm i} = 3.4\times 10^{-8}M_{\rm Pl}$ (left panels) and $\chi_{\rm i}=1.0\times 10^{-8}M_{\rm Pl}$ (right panels) which we already chose for Fig~\ref{fig:GWspectra}. The snapshots are taken at times during the non-linear evolution of the fields, in $\Delta\tau = 2$ intervals between $\tau = 73$ and $\tau = 79$, just when the GW production is strongest. 
Fig.~\ref{fig:GWspectra} and Fig.~\ref{fig:2d_snapshots} together demonstrate very clearly that there is a correlation between the gradients of $\chi$ and the amplitude of the produced GWs: for $\chi_{\rm i} = 3.4\times 10^{-8}M_{\rm Pl}$, the gradients and, consequently the GW amplitude, are higher than for $\chi_{\rm i} = 1.0\times 10^{-8}M_{\rm Pl}$.
We postpone an analysis of the physical reason for this sensitive dependence of the gradients of $\chi$ (and therefore of the GWs) on the initial value $\chi_{\rm i}$ to Section~\ref{dynamics}. 

\subsection{Toolkit for computing GW Anisotropies}
\label{subsec:Anisotropies}

The amount of GW production depends strongly on the value of $\chi_{\rm i}$, as we have shown explicitly in Section~\ref{subsec:GW_vs_InitialChi} for two values of $\chi_{\rm i}$. We will later present the data from many simulations, each with a different $\chi_{\rm i}$ amplitude drawn from the appropriate random distribution. We will find that in the scenario we are studying, the dependence of $\Omega_\GW$ on $\chi_{\rm i}$ is very irregular (see Fig.~\ref{chiGWtot}). To quantify the amount of anisotropy as derived from a given $\Omega_\GW(\chi_{\rm i})$ dependence, we first need to provide a mathematical toolkit for such an analysis.

To begin with, let us assume a situation where $\Omega_\GW(\chi_{\rm i})$ depends linearly on $\chi_{\rm i}$. We will not need this to be the case in general (and actually it is not), but it will be instructive to study the linear relation as a starting point. Normalizing the $\chi_{\rm i}$ variations to the natural scale of the problem, $H_*$, we can then write
\be
\label{equ:lindep}
\Omega_\GW(\chi_{\rm i}) = c_0 + c _1{\delta\chi_{\rm i}\over H_*}\,,
\ee
with $\delta\chi_{\rm i} \equiv \chi_{\rm i}-\overline{\chi}_{\rm i}$, where $\overline{\chi}_{\rm i}$ is the mean value over the currently observable universe. The constants $c_0, c_1$ are dimensionless and completely characterize the function $\Omega_\GW(\chi_{\rm i})$ (under the linear assumption). From Eq.~(\ref{equ:lindep}) one can easily see that $c_0$ can be identified with the mean amplitude of the GWs over the observable universe, i.e.~$c_0 \equiv \overline{\Omega}_\GW$. We can then express the relative fluctuations of the GW energy density as
\begin{equation}\label{eq:deltaGW}
\delta\Omega_\GW \equiv {\Omega_\GW - \overline{\Omega}_\GW\over\overline{\Omega}_\GW} \equiv {c_1\over c_0}{\delta\chi_{\rm i}\over H_*} \, .
\end{equation}
As these fluctuations are proportional to $\delta\chi_{\rm i}$, like $\chi_{\rm i}$ they represent a nearly Gaussian and scale-invariant random field. The power spectrum of $\delta\Omega_\GW$ can then be directly related to the power spectrum ${\cal P}_{\chi}$ of $\chi_{\rm i}$, simply as
\begin{eqnarray}
{\cal P}_{\GW} = {c_1^2\over c_0^2}{{\cal P}_{\chi}\over H_*^2} = {1\over 4\pi^2} {c_1^2\over c_0^2} \, .
\end{eqnarray}
To measure fluctuations on the celestial sphere, it is better to express them in terms of spherical harmonics $\lbrace {\rm Y}_{lm}\rbrace$. This makes it possible to characterise the statistical properties of $\delta\Omega_\GW$ in terms of an angular power spectrum, in the same way as one does for the CMB temperature anisotropies. We can decompose the fluctuations in the GW energy density as
\begin{eqnarray}
\delta\Omega_\GW(\hat n) = \sum_{l \geq 1}^{\infty} \sum_{m = -l}^{+l} g_{lm}{\rm Y}_{lm}(\hat n)\,,
\end{eqnarray}
where $g_{lm} = \int_{4\pi}d\Omega {\rm Y}^*_{lm}(\hat n)\delta\Omega_\GW(\hat n)$ are (complex) coefficients weighting each angular moment. The angular power spectrum $C_l$ is then defined as the ensemble average of such coefficients,
\begin{eqnarray}\label{eq:C_l}
\left\langle \,g^*_{lm}g_{l'm'}\, \right\rangle \equiv C_l\delta_{ll'}\delta_{mm'}\,,
\end{eqnarray}
where the Kronecker delta $\delta_{ll'}$ reflects statistical isotropy whereas $\delta_{mm'}$ shows the statistical independence of the $2l+1$ multipoles for a given angular mode $l$. The $C_l$'s are given by
\begin{eqnarray}\label{eq:C_lv2}
C_l \equiv 2\pi \int d\cos\theta P_l(\cos\theta) C(\cos\theta)\,,
\end{eqnarray}
where $P_l(\cos\theta)$ are the Legendre polynomials, and $C(\cos\theta)$ is the angular correlation of the GW fluctuations at different directions in the sky $\hat n_1$ and $\hat n_2$:
\begin{eqnarray}
C(\cos\theta) \equiv \left\langle \delta\Omega_\GW(\hat n_1)\delta\Omega_\GW(\hat n_2) \right\rangle\,,
\end{eqnarray}
with $\hat n_1\hspace{-0.6mm}\cdot\hspace{-0.2mm}\hat n_2 \equiv \cos\theta$.

Equivalently, the angular correlation can be expressed as a linear sum in the $C_l$'s weighted as
\begin{eqnarray}
\left\langle \delta\Omega_\GW(\hat n_1)\delta\Omega_\GW(\hat n_2) \right\rangle = \sum_{l \geq 1}^{\infty} {(2l+1)\over 4\pi}\,C_l\,P_l(\hat n_1\hspace{-0.6mm}\cdot\hspace{-0.2mm}\hat n_2)\,.\nonumber\\
\end{eqnarray}
Thanks to the assumed linear relation between $\delta\Omega_\GW$ and $\delta\chi_{\rm i}$ in Eq.~(\ref{eq:deltaGW}), the angular power spectrum of the GW energy density fluctuations can then be calculated  very easily. Indeed, the calculation is analogous to the Sachs-Wolfe plateau~\cite{LiddleLyth} for temperature fluctuations on large scales (i.e.~small multipole $l$) due to scalar perturbations. It is simply given by
\be
\label{equ:Cl}
l(l+1)C_l=\frac{\pi}{2}{\cal P}_{\GW}=\frac{1}{8\pi}{c_1^2\over c_0^2} \, .
\ee
Therefore, as long as $\delta\Omega_\GW$ is linearly dependent on $\delta\chi_{\rm i}$ as in Eq.~(\ref{equ:lindep}), the coefficients $c_0$ and $c_1$ completely determine the angular power spectrum. In the case of massless preheating, and generally in any other scenario of preheating, the $\Omega_\GW(\chi_{\rm i})$ relationship will not be linear. This problem was recently discussed in Ref.~\cite{Suyama:2013dqa}, which motivated our approach.

To describe fluctuations on any angular scale independent of the functional form of the relation $\Omega_\GW(\chi_{\rm i})$, we need to compute the two-point correlation function of the GW energy density originating from two points $\bx$ and ${\bf y}$. Due to isotropy, this correlator can only depend on the separation $|\vx-\vy|$. It can be written as
\bea\label{eq:GWcorr}
&&\left\langle \Omega_{\GW}(\bx)\Omega_{\GW}({\bf y}) \right\rangle \equiv \nonumber\\ 
&&~~~~~\int d\chi_{\bf x}d\chi_{\bf y}P(\chi_{\bf x},\chi_{\bf y}) \Omega_{\GW}(\chi_{\bf x})\Omega_{\GW}(\chi_{\bf y})\,,
\eea
where $P(\chi_{\bf x},\chi_{\bf y})$ is the joint probability distribution for the field values $\chi_{\bf x}=\chi_{\rm i}({\bf x})$ and $\chi_{\bf y}=\chi_{\rm i}({\bf y})$ at the points $\vx$ and $\vy$. Defining the vector $\vec{\delta\chi} \equiv (\chi_{\bf x}-\overline\chi_{\rm i},\chi_{\bf y}-\overline\chi_{\rm i})$, since these are Gaussian random fields, we have
\be
P(\chi_{\bf x},\chi_{\bf y}) = {1\over 2\pi\sqrt{|G|}}\,e^{-{1\over2}\vec{\delta\chi}^{\rm T}G^{-1}\vec{\delta\chi}}\,,
\ee
where the $2\times2$ covariant matrix $G$ and its inverse $G^{-1}$, with determinant $|G|$, are given by
\begin{eqnarray}
G \equiv 
\left(\hspace{-1mm}
\begin{array}{cc}
G_{{\bf x},{\bf x}} & G_{{\bf x},{\bf y}}   \\
G_{{\bf x},{\bf y}} & G_{{\bf y},{\bf y}}
\end{array}
\hspace{-1mm}\right)\,, ~~G^{-1} \equiv 
{1\over|G|}\hspace{-1mm}\left(\hspace{-1mm}
\begin{array}{cc}
G_{{\bf y},{\bf y}} & -G_{{\bf x},{\bf y}}   \\
-G_{{\bf x},{\bf y}} & G_{{\bf x},{\bf x}}
\end{array}
\hspace{-1mm}\right)\nonumber\\
\end{eqnarray}
with $G_{{\bf x},{\bf y}}\equiv\langle\delta\chi_{\rm i}({\bf x})\delta\chi_{\rm i}({\bf y})\rangle$ the field correlator,
and $\sigma_\chi^2 = G_{{\bf x},{\bf x}}=\langle\delta\chi^2\rangle$ the field variance [see Eq.~(\ref{chivar})].  From the scale-invariant power spectrum (\ref{eq:ChiSpectrum}) we find
\be
G_{{\bf x},{\bf y}}\approx \frac{H_*^2}{4\pi^2}\ln(|{\bf x}-{\bf y}|H_0).
\ee

\begin{figure}[t]
\begin{center}
\includegraphics[width=8.0cm]{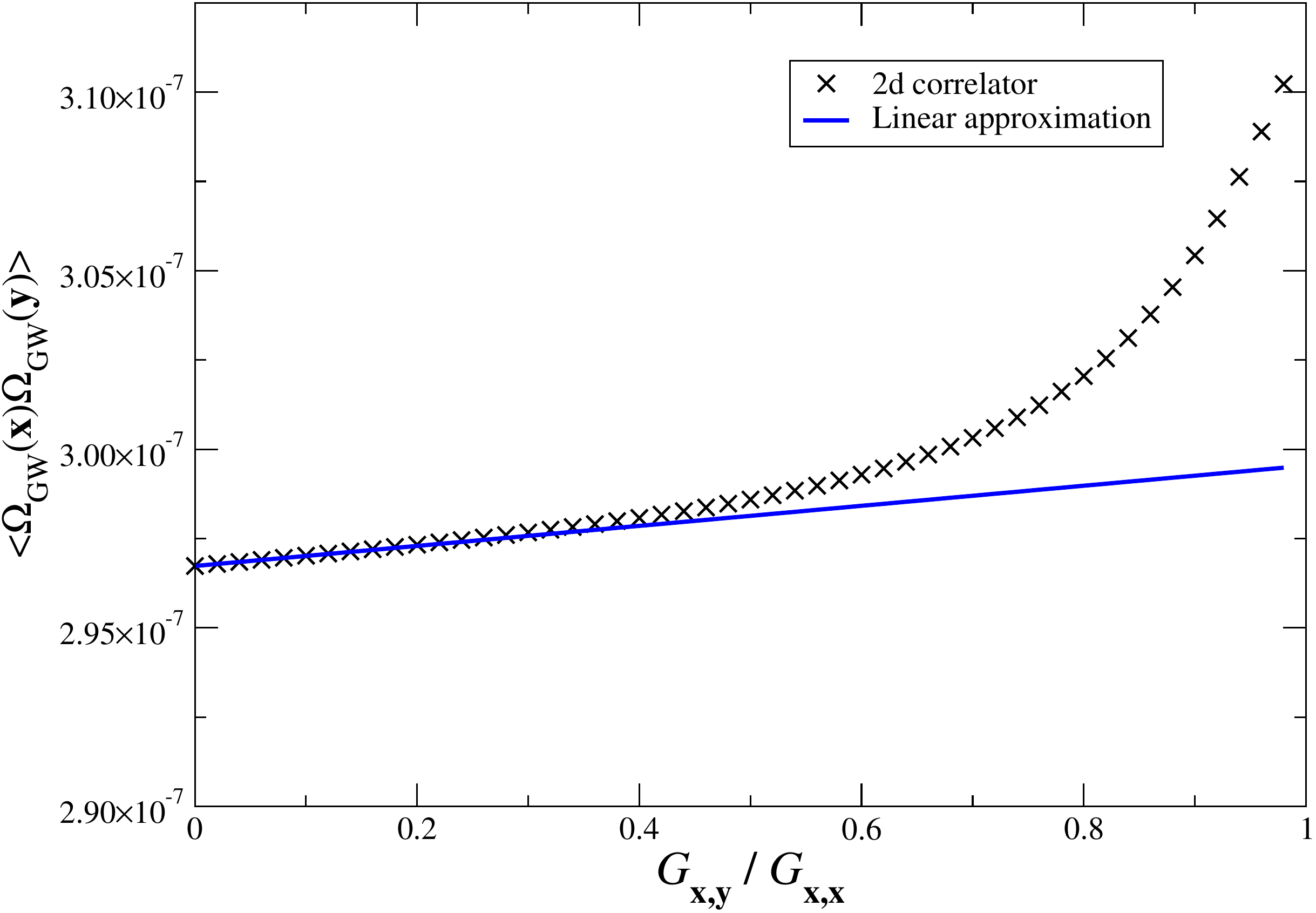}
\end{center}
\caption{The full GW energy density correlator and its linearised version. The two results agree very well up to the largest currently observable scales, i.e.~for small $G_{{\bf x},{\bf y}}/\sigma_\chi^2$ values.}
\label{corr2d}
\end{figure}

By obtaining the function $\Omega_\GW(\chi_{\rm i})$ from lattice simulations, we can compute the GW energy density correlator (\ref{eq:GWcorr}). This is shown in Fig.~\ref{corr2d} for $\chi_{\rm i}=3.42\times 10^{-7}M_{\rm Pl}$. Note that the correlator only depends on the distance $|{\bf x}-{\bf y}|$ through the ratio $G_{{\bf x},{\bf y}}/G_{{\bf x},{\bf x}}=G_{{\bf x},{\bf y}}/\sigma_\chi^2$.
In principle one can use this to compute 
angular correlation of the GW energy density at two directions $\hat n_1, \hat n_2$ in the sky, by evaluating Eq.~(\ref{eq:GWcorr}) at positions $\vx = R\hat n_1$ and $\vy = R\hat n_2$, with $R$ ($\sim H_o^{-1}$) the distance to the 'scattering surface' at preheating where the GWs were emitted. From there we can obtain the angular power spectrum $C_l$ by means of Eq.~(\ref{eq:C_lv2}). 

In practice, this procedure can be cumbersome and, more importantly, since $\Omega_\GW(\chi_{\rm i})$ may be very irregular, it would be difficult to assess the accuracy in the final amplitude of the $C_l$'s. 
Instead, we can note that 
for observationally relevant scales, $|{\bf x}-{\bf y}|\sim 1/H_0$, the ratio
\be {G_{{\bf x},{\bf y}}\over\sigma_\chi^2} \approx {1\over N_*} \simeq 0.017 \ee
is small. As Fig.~\ref{corr2d} shows, the correlator is very well approximated by a linear Taylor expansion approximation, which we discuss next.

To simplify the analysis and to avoid having to compute the full correlation function Eq.~(\ref{eq:GWcorr}), we  
perform a linear Taylor expansion of the joint probability distribution in powers of the field correlator normalized to the variance, $G_{{\bf x},{\bf y}}/\sigma_\chi^2$, as
\begin{eqnarray}\label{eq:TaylorExp}
P(\chi_{\bf x},\chi_{\bf y}) \propto \exp\left[-\frac{(\delta\chi_{\bf x}^2+\delta\chi_{\bf y}^2)-2\left(G_{{\bf x},{\bf y}}\over\sigma_\chi^2\right)\delta\chi_{\bf x}\delta\chi_{\bf y}}{2\sigma_\chi^2\left[1-\left(G_{{\bf x},{\bf y}}\over\sigma_\chi^2\right)^2\right]}\right]\nn\\
\approx \exp\left(-\frac{\delta\chi_{\bf x}^2}{2\sigma_\chi^2}\right)\exp\left(-\frac{\delta\chi_{\bf y}^2}{2\sigma_\chi^2}\right)\left[1+\left(G_{{\bf x},{\bf y}}\over\sigma_\chi^2\right)\delta\chi_{\bf x}\delta\chi_{\bf y} \right. \nonumber\\ 
\left. +~ \mathcal{O}\hspace{-.5mm}\left(G_{{\bf x},{\bf y}}\over\sigma_\chi^2\right)^2\right]\nonumber
\end{eqnarray}
Substituting this expansion into Eq.~(\ref{eq:GWcorr}), we then obtain
\begin{eqnarray}
\label{equ:linexp}
\left\langle \Omega_{\GW}({\bf x})\Omega_{\GW}({\bf y}) \right\rangle \simeq\hspace{5cm} \\
\langle\Omega_\GW(\chi_{\rm i})\rangle^2+\frac{\left\langle
\delta\chi_{\rm i}\Omega_\GW(\chi_{\rm i})
\right\rangle^2
}{\sigma_\chi^2}
\left(G_{{\bf x},{\bf y}}\over\sigma_\chi^2\right) + ~\mathcal{O}\hspace{-.5mm}\left(G_{{\bf x},{\bf y}}\over\sigma_\chi^2\right)^2,\nonumber
\end{eqnarray}
where the expectation values on the right hand side are given by
\begin{eqnarray}\label{eq:GWav}
\langle\Omega_\GW\rangle  &\equiv& \int d\chi_{\rm i} P(\chi_{\rm i}) \Omega_\GW(\chi_{\rm i})\,\\
\left\langle
\delta\chi_{\rm i}\Omega_\GW(\chi_{\rm i})
\right\rangle &\equiv& \int d\chi_{\rm i} P(\chi_{\rm i}) \delta\chi_{\rm i}\Omega_\GW(\chi_{\rm i})\,
\end{eqnarray}
computed with the single-point probability distribution
\be
\label{equ:singleP}
P(\chi_{\rm i}) = {1\over\sqrt{2\pi}\,\sigma_\chi}\exp\left\{-{1\over2}{(\chi_{\rm i}-\overline{\chi}_{\rm i})^2\over\sigma_\chi^2}\right\}. 
\ee

Re-arranging the terms on the right-hand side of Eq.~(\ref{equ:linexp}), we 
can write the equation as
\begin{eqnarray}
\label{equ:linexpV2}
\left\langle \Omega_{\GW}({\bf x})\Omega_{\GW}({\bf y}) \right\rangle \simeq \hspace{5cm} \\
\left\langle \left( \langle\Omega_\GW\rangle + {\left\langle
\delta\chi_{\rm i}\Omega_\GW(\chi_{\rm i})
\right\rangle\over\sigma_\chi^2}\delta\chi_{\rm i}(\vx)\right)\hspace{2cm}\right.\nonumber\\
\times \left.\left( \langle\Omega_\GW\rangle + {\left\langle
\delta\chi_{\rm i}\Omega_\GW(\chi_{\rm i})
\right\rangle\over\sigma_\chi^2}\delta\chi_{\rm i}(\vy)\right)\right\rangle\nonumber
\end{eqnarray}
This is precisely the form of Eq.~(\ref{equ:lindep}), and therefore we conclude that the results obtained earlier in the linear case can be applied generally, provided that we identify the coeffocients as
\begin{eqnarray}\label{eq:GWcoeffs}
c_0 = \langle\Omega_\GW\rangle\,,~~~~~ c_1 = {H_*\over\sigma_\chi^2}\left\langle
\delta\chi_{\rm i}\Omega_\GW(\chi_{\rm i})
\right\rangle.
\end{eqnarray}
We can therefore use Eq.~(\ref{equ:Cl}) directly to compute the angular power spectrum of the relative GW energy density fluctuations. We obtain
\be
\label{equ:Cl2}
l(l+1)C_l=\frac{H_*^2}{8\pi}\frac{\langle\delta\chi_{\rm i}\Omega_\GW(\chi_{\rm i})\rangle^2}{\sigma_\chi^4\langle\Omega_\GW\rangle^2}.
\ee
This equation is one of the main results of the paper. It is a master formula for the angular power spectrum of the energy density fluctuations of any GW background of cosmological origin, whose anisotropies originated from the modulation due to an inflationary spectator field. Contrary to the temperature fluctuations in the CMB, which dynamically evolve during the tight coupling matter-radiation era, GWs decouple upon production. Thus, the GW energy density angular power spectrum will have the characteristic 'plateau' shape $l(l+1)C_l = $ const for every multipole $l$. Therefore only very small multipoles (large angular scales) will matter, since for large multipoles (small angular scales) the spectrum will decay as $C_l \sim 1/l^2$. If detectable, the effect would probably be easiest to measure on the level of the quadrupole ($l = 2$).

We are now ready to calculate the typical amplitude of fluctuations $\sim\sqrt{l(l+1)C_l}$ for any value of $\overline\chi_{\rm i}$, by simply evaluating the expectation values in Eq.~(\ref{equ:Cl2}) from the results of our lattice simulations numerically. We will now apply our anisotropies machinery to the specific case of the GW background from massless preheating.

\subsection{The anisotropies in the GW background from massless preheating}
\label{subsec:Anisotropies2}

\begin{figure}[t]
\begin{center}
\includegraphics[width=8.0cm]{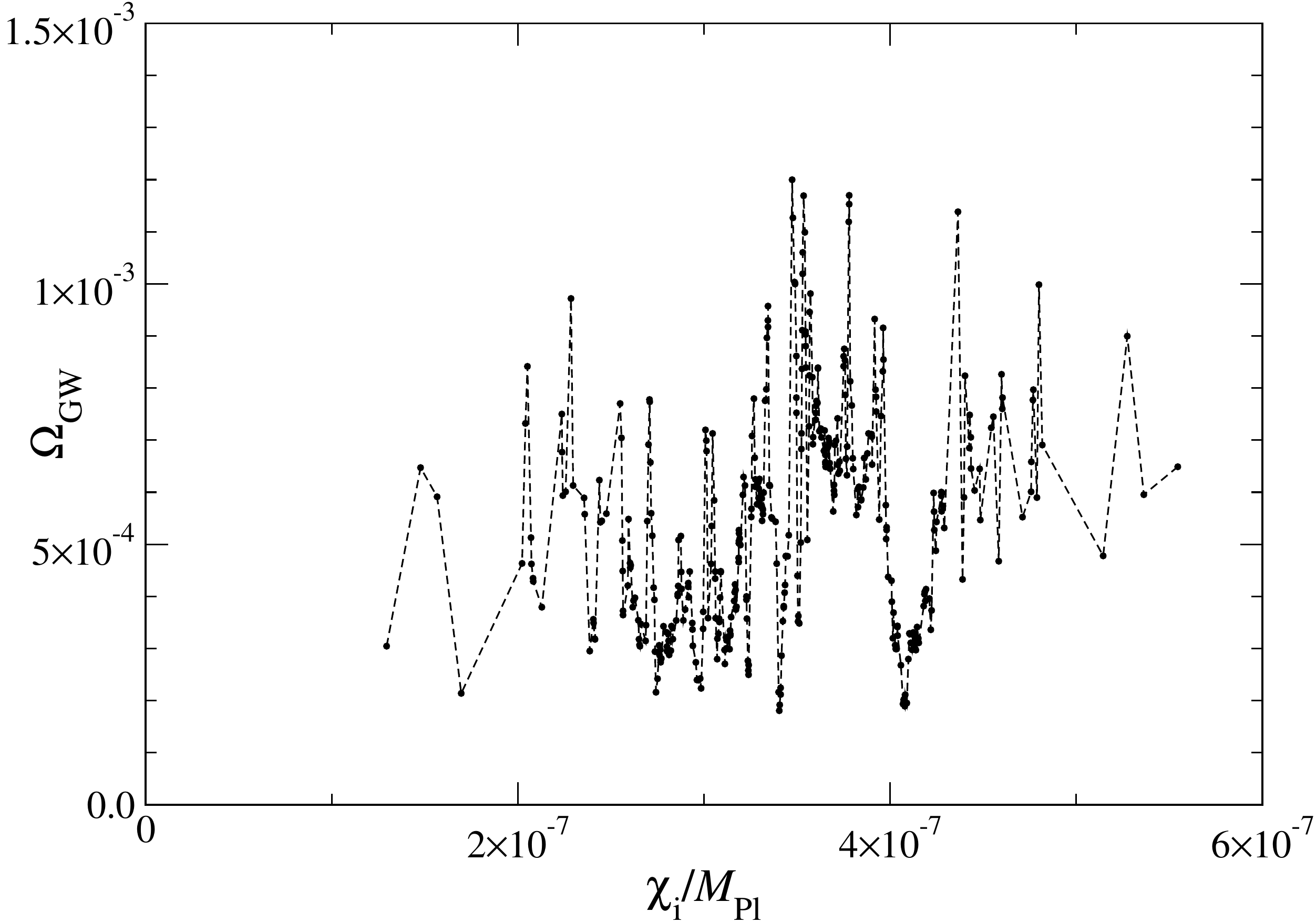}
\end{center}
\caption{$\Omega_\GW$ for our sample of initial field values $\chi_{\rm i}$. The straight lines inbetween the points (real data from the smulation) are drawn to guide the eye, but in reality more random oscillations are expected in between points.}
\label{fig:GWvsChi}
\end{figure}

According to the considerations in section \ref{sims}, we chose a mean value $\overline{\chi}_{\rm i}=3.42 \times 10^{-7}M_{\rm Pl}$ and a variance $\sigma_\chi^2=3.3\times 10^{-15}M_{\rm Pl}^2$ to describe the range of initial $\chi_{\rm i}$ values the GW background from preheating is likely to have originated from. We used the Monte Carlo method to randomly choose ${\cal N}= 400$ random values $\chi_{\rm i}^j$, $j\in\{1,\ldots,{\cal N}\}$ 
from the Gaussian distribution (\ref{equ:singleP}). To be exact, we randomly picked 200 values $\chi_{\rm i}^j$ and chose the remaining half to be the symmetric value $\chi_{\rm i}^{j'}=2\overline{\chi}_{\rm i}-\chi_{\rm i}^j$. This ensures that the mean of our distribution will be exactly the required $\overline{\chi}_{\rm i}=3.42 \times 10^{-7}M_{\rm Pl}$, which reduces the error when computing expectation values. This is necessary as we only have a finite sample of values so will will never be able to obtain a perfect Gaussian distribution.

The Monte Carlo method has the advantage of making it easier to sample the highly chaotic variation of the GW energy density $\Omega_\GW(\chi_{\rm i}^j)$ without needing to use a very small step size in $\chi$, as well as simplifying the computation of the expectation values  in Eq.~(\ref{equ:Cl2}).
For each $\chi_{\rm i}^j$, we did one simulation run, measuring the GW energy density $\Omega_\GW(\chi_{\rm i}^j)$, see Fig.~\ref{fig:GWvsChi}.
As the plot shows, $\Omega_\GW$
is highly dependent on $\chi_{\rm i}$, varying by as much as a factor of five between nearby values, although there are some ranges of $\chi_{\rm i}$ where the dependence is much smoother. This irregular behavior is in line with the chaotic dynamics observed earlier~\cite{Tanaka:2003cka,Suyama:2006rk,Bond}, but its amplitude is unexpectedly high. We also measured the variation of $\Omega_\GW$ between different random realisations of the field fluctuations and found that it was much smaller than the variation between different values of $\chi_{\rm i}$, which indicates that the effect is not merely statistical fluctuation.

As we used a Monte Carlo method to choose our range of $\chi_{\rm i}^j$, the expectation values in Eq.~(\ref{equ:Cl2}) can be approximated by averages within our sample, 
\bea
\label{equ:averages}
\langle\Omega_\GW\rangle&\approx& \frac{1}{\cal N}\sum_j\Omega_\GW(\chi_{\rm i}^j),
\nonumber\\
\langle\delta\chi\Omega_\GW\rangle&\approx& \frac{1}{\cal N}\sum_j(\chi_{\rm i}^j-\overline{\chi}_{\rm i})\Omega_\GW(\chi_{\rm i}^j).
\eea
For $\overline\chi_{\rm i}=3.42\times 10^{-7}M_{\rm Pl}$, we obtained
$\langle\Omega_\GW\rangle= (5.45\pm 0.13)\times 10^{-4}$ and $\langle\delta\chi\Omega_\GW\rangle= (3.0\pm1.2)\times 10^{-12}M_{\rm Pl}$. Substituting these into Eq.~(\ref{equ:Cl2}) gives the amplitude of the relative fluctuations $\delta\Omega_\GW=(\Omega_{\GW}/\overline{\Omega}_\GW-1)$ as 
\be
\sqrt{l(l+1)C_l}=0.017\pm0.003,
\ee
where the errors are estimated by the bootstrap method. This method provides a useful way of measuring the uncertainty in expectation values calculated from a set of data, by mimicking the process of obtaining new data from the same probability distribution. Assuming there are $N$ data points in the original ensemble, for each bootstrap sample $N$ of these points are randomly selected, without avoiding double counting. The expectation value is then calculated based on the current set of data points, and the variance of many such bootstrap samples gives an estimate of the error in the expectation value. 
In our case, we used 1000 bootstrap samples of 200 randomly chosen symmetric pairs $\chi_{\rm i}^{j'}$,$\chi_{\rm i}^j$ (again to make sure that each bootstrap sample has the correct mean $\overline{\chi}_{\rm i}=3.42 \times 10^{-7}M_{\rm Pl}$) to calculate (\ref{equ:averages}), and the variance of these samples gave an error estimate of magnitude $0.003$ for the amplitude of relative fluctuations.

The numerical value of the variance for our set of initial $\chi_{\rm i}$, generated by the Python random number generator, turned out to be $\sigma^2=4.3\times 10^{-15}M_{\rm Pl}^2$, significantly higher than the desired variance $\sigma_\chi^2=3.3\times 10^{-15}M_{\rm Pl}^2$. 
To rectify this, we 
reweighted our data to resemble a sample with a variance closer to the required one.

Reweighting makes it possible to use to Monte Carlo data for a specific probability distribution to calculate expectation values for other, similar distributions. Assume values $x$ were drawn from a probability distribution $p(x)$ and you need to calculate the expectation value of an observable $O$ from a slightly different probability distribution function $p'(x)$:
\be \langle O \rangle' = \frac{\int dx p'(x) O(x)}{\int dx p'(x)} \ee
We can re-express this in terms of the old probability distribution $p(x)$ as
\begin{eqnarray}
 \langle O \rangle' &=& \frac{\int dx p(x) \frac{p'(x)}{p(x)} O(x)}{\int dx p(x) \frac{p'(x)}{p(x)}}
\nonumber\\
&=& \frac{\langle r(x) O(x) \rangle}{\langle r(x) \rangle}\equiv \frac{\sum_j r(x_j) O(x_j)}{\sum_j r(x_j)} 
\end{eqnarray}
where  $r(x)=\frac{p'(x)}{p(x)}$ is the reweighting factor. Therefore, to calculate expectation values from a slightly different probability distribution to the original one, we can simply reweight each observable by $r(x_j)$. As long as the probability distributions are close to each other, i.e. $\frac{1}{N}\sum_j r(x_j) \approx 1$, this method can be trusted.

In our case, the numerical data suggested a Gaussian probability distribution with mean $\overline{\chi}_{\rm i}$ and variance $\sigma^2$ which we need to reweight to have the correct variance $\sigma_\chi^2=3.3\times 10^{-15}M_{\rm Pl}^2$, i.e.
\be p'(\chi_{\rm i})=\frac{1}{\sqrt{2\pi \sigma_{\chi}^2}}\exp\left(-\frac{(\chi_{\rm i}-\overline{\chi}_{\rm i})^2}{2\sigma_\chi^2}\right)\ee  
Note that the reweighting takes place for the whole sample (when calculating the mean expectation value) and for each bootstrap sample (when estimating the errors).
\begin{figure}[t]
\begin{center}
\includegraphics[width=8.0cm]{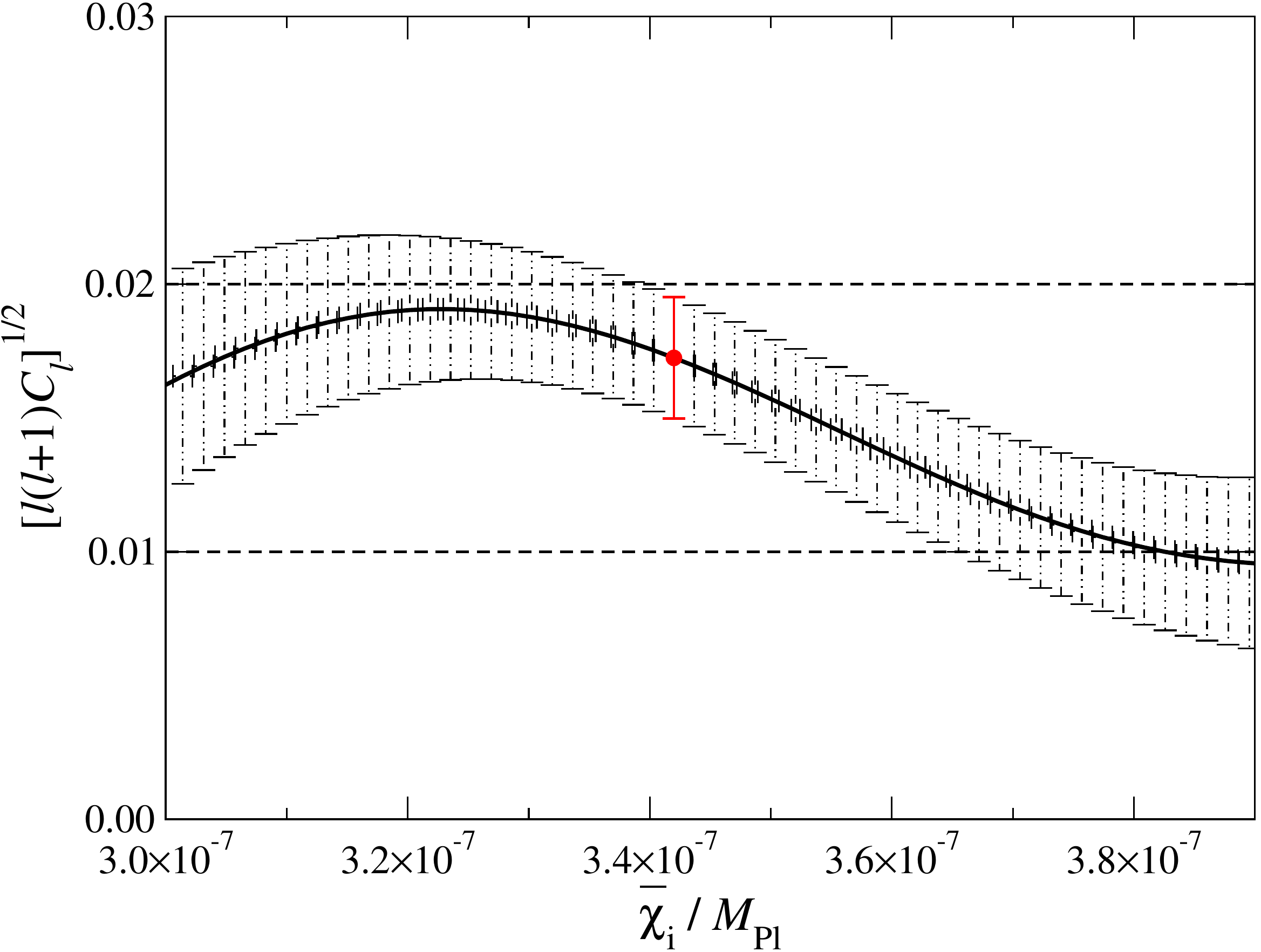}
\end{center}
\caption{
\label{fig:Clvschi}
The relative amplitude of the multipoles of the GW background as a function of the average field value $\overline{\chi}_{\rm i}$, calculated from Eq.~(\ref{equ:Cl2}). The red dot shows the amplitude for original mean value $\overline{\chi}_{\rm i}=3.42\times 10^{-7}M_{\rm Pl}$, and the curve shows values obtained by reweighting the same data.
}
\end{figure}

By employing the method of reweighting, we can also use our Monte Carlo data to calculate expectation values around different nearby mean values $\overline{\chi}_{\rm i}'$ (i.e. a $\chi$ background with a slightly different average across our observable Universe). These will correspond to a probability distribution
\be p'(\chi_{\rm i})=\frac{1}{\sqrt{2\pi \sigma_{\chi}^2}}\exp\left(-\frac{(\chi_{\rm i}-\overline{\chi}'_{\rm i})^2}{2\sigma_\chi^2}\right)\ee  
where $\overline{\chi}'_{\rm i}$ is a different mean value to our chosen one. The total reweight factor is therefore
\be
r(\chi_{\rm i}^j)=\frac{\sigma}{\sigma_{\chi}}\exp\left[-\frac{(\chi_{\rm i}^j-\overline\chi_{\rm i}')^2}{2\sigma_\chi^2}
+\frac{(\chi_{\rm i}^j-\overline\chi_{\rm i})^2}{2\sigma^2}\right]
\ee 
We can use this procedure to calculate $C_l$ from the expectation values in (\ref{equ:averages}), evaluated around the new probability distribution. The solid line in Fig.~\ref{fig:Clvschi} shows the relative amplitude of angular fluctuations for different mean values $\overline{\chi}_{\rm i}$ across our observable Universe, where the red dot corresponds to our original choice $\overline{\chi}_{\rm i}=3.42 \times 10^{-7}M_{\rm Pl}$. For the reweighted mean values, the error bars in the fluctuations have been obtained by the bootstrap method, similarly to the original value. 

One point to note about the plot is that the reweighted data sets have an uncertainty in the value of $\overline{\chi}_{\rm i}'$, because once reweighted each bootstrap sample (which is chosen to have a mean $\overline{\chi}_{\rm i}=3.42 \times 10^{-7}M_{\rm Pl}$) has a slighlty different mean value.
The uncertainty in $\overline{\chi}_{\rm i}'$ becomes larger far away from the original mean, where we do not have enough coverage to simulate a probability distribution with the chosen new mean. This also implies that $\overline{\chi}_{\rm i}$ in equation (\ref{equ:averages}) needs to be replaced by its reweighted value over the chosen probability distribution, for the whole sample and each bootstrap sample separately. 

Note also that in a previous letter about this work \cite{LauraDaniArttuPRL}, we neither reweighted the mean values nor did we choose symmetric pairs when carrying out the bootstrap error analysis. This led to a large over-estimation of the error (see Fig. 3 in \cite{LauraDaniArttuPRL}), as the uncertainties in the different expectation values were correlated. We believe that the error analysis performed for the current paper is more representative of the real uncertainty, which comes about mainly due to the irregular behaviour of $\Omega_\GW(\chi_{\rm i}^j)$.

For most of the range of $\overline\chi_{\rm i}$ presented in Fig. \ref{fig:Clvschi}, the amplitude of the fluctuations is above the one percent level, even within error bars. This is much higher than the relative amplitude of fluctuations in the CMB which is of order $10^{-5}$. It is reasonable to hope that fluctuations of the order of $1\%$ could be measured by future GW detectors, although it is difficult to make any statement about their sensitivity to anisotropies at the current stage.

\begin{figure}[t]
\begin{center}
\includegraphics[width=8.0cm]{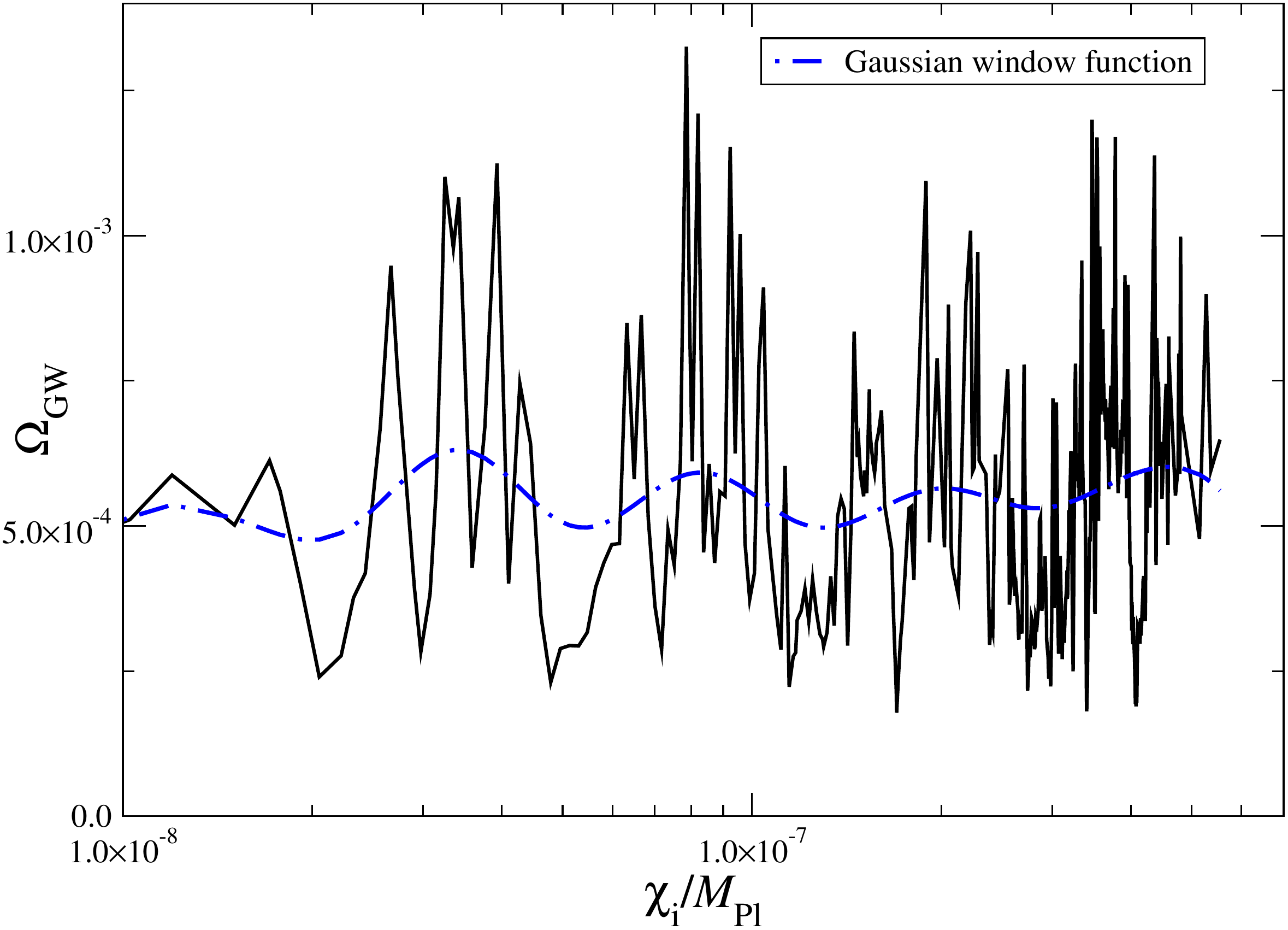}
\end{center}
\caption{$\Omega_\GW$ for a larger range of field values $\chi_{\rm i}$. The solid curve corresponds to the convolution of the data with a Gaussian window function.}
\label{chiGWtot}
\end{figure}

As described in section \ref{sepuni}, the mean value of $\chi$ across our observable universe, $\overline\chi_{\rm i}$, is a free parameter dependent on the total number of e-folds of inflation. To have a complete picture of the anisotropies in the GW background, we should therefore analyse a wider range of $\chi_{\rm i}$ values than the one presented so far. 
The GW energy density for a larger range of $\chi_{\rm i}$ (including as the largest values the results from Fig.~\ref{fig:GWvsChi}) is presented in Fig. \ref{chiGWtot} on a logarithmic scale. As it is reasonable to assume inflation lasted some number of e-folds longer than the minimal required number of $N_*=50$, very small values of $\chi_{\rm i}$ are  unlikely (as even with only $N_{\rm tot}=60$ we would have at least an expected value of order $\overline\chi_{\rm i}  \sim 3 \times 10^{-8}M_{\rm Pl}$). Correspondingly, $\chi_{\rm i}$ can be larger if inflation lasted for a very long time.

The data presented in Fig.~\ref{chiGWtot} are mainly for illustration, because obtaining statistically significant results would require more data. 
However, the plot also reveals some non-trivial structure. In particular, the GW energy density has an approximate log-periodic dependence on $\chi_{\rm i}$, 
with regions of high, quickly varying GW amplitude alternating with regions of low amplitude. To make this more apparent, we have also included a curve that shows the convolution of the data with a Gaussian window function,
\be \widetilde{\Omega}_\GW(\log \chi) = \frac{1}{\sqrt{2\pi \sigma_{\rm w}^2}} \int d\delta\, e^{-\delta^2/2\sigma_{\rm w}^2} \Omega_\GW(\log \chi+\delta) \, , \ee
where $\sigma_{\rm w}^2=0.05$ is the spread of the window function. This log-periodic structure was predicted by \cite{Bond} and will be explained in the next section. 

\subsection{Field dynamics}\label{dynamics}

\begin{figure}[t]
\begin{center}
\includegraphics[width=8.0cm]{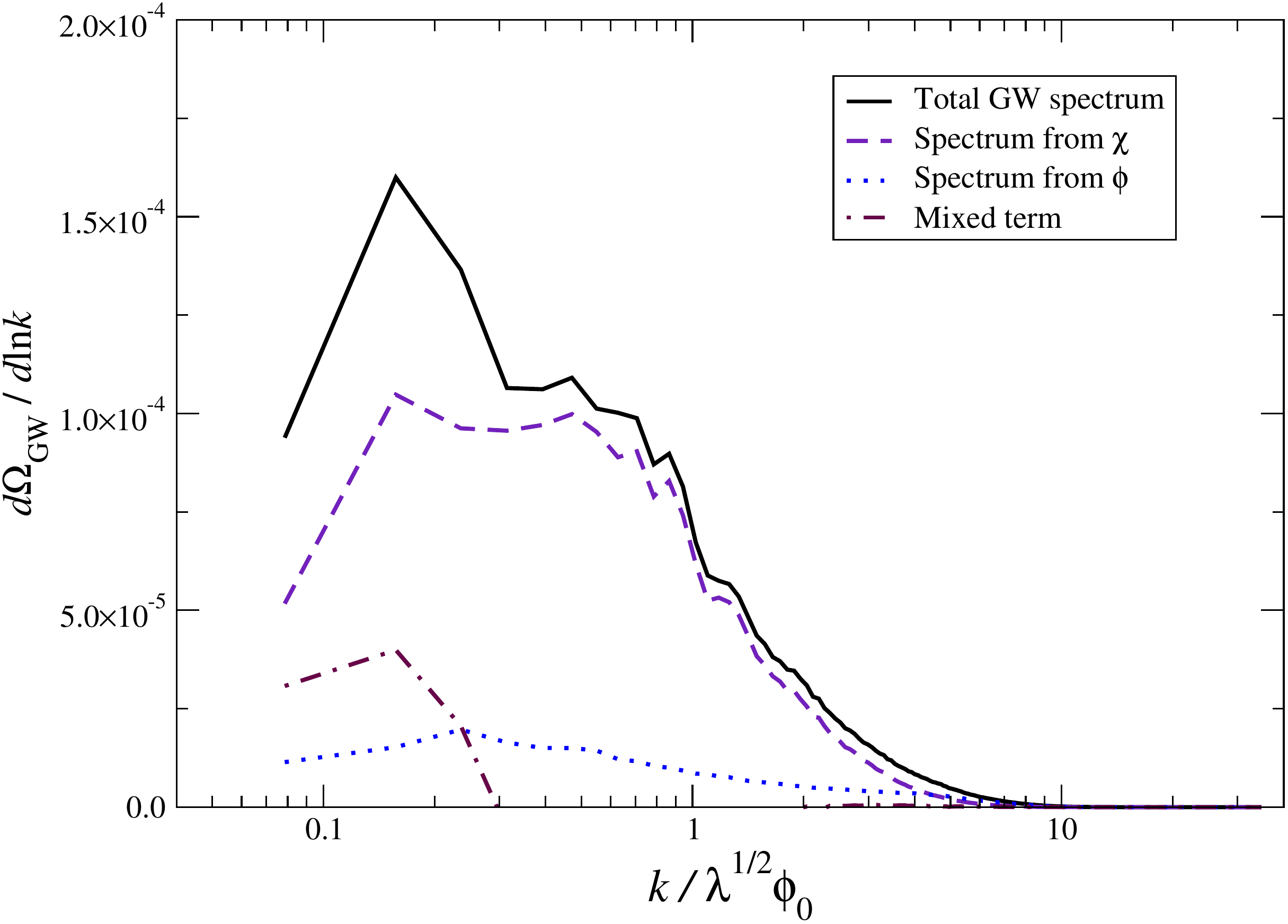}
\end{center}
\caption{The GW amplitude from two different $\chi_{\rm i}$ sourced by only $\phi$, $\chi$ and both fields respectively}
\label{GWsources}
\end{figure}

In order to further understand the physical origin of the sensitive dependence of the GW amplitude on $\chi_{\rm i}$, here we study the relationship between GW production and the field dynamics. As the source term for tensor perturbations is given by the field gradients (and will appear quartically in the equation for the GW power spectrum (\ref{GWspect})), it is natural to ask which of the scalar fields is primarily responsible for the production of GWs. In Fig.~\ref{GWsources} we have plotted the same power spectrum as in Fig.~\ref{fig:GWspectra}, however additionally we plotted spectra obtained from using only $\phi$ or $\chi$ as a source of GWs (the total GW amplitude will also contain cross terms between the fields). 
The plot shows that the GWs are sourced primarily by the gradients of the $\chi$ field, and therefore we can focus on its dynamics in order to understand the physical origin of the variation of the GW energy density.

In Ref.~\cite{Bond} it was observed that the evolution of the system strongly depends on the relative phase of the homogeneous modes $\phi(t)$ and $\chi(t)$ at the time the field dynamics become non-linear (i.e. when $\chi$ becomes sufficiently large). In particular, in some cases $\chi(t)$ acquires a very large amplitude compared to the inflaton, leading to curvature spikes. If the inflaton oscillates (during the linear stage) with period $T$, initial $\chi$ configurations related by 
\begin{eqnarray}\label{eq:ChiToChi}
{\chi_{\rm i}'\over\chi_{\rm i}} = e^{\mu n T},
\end{eqnarray}
with $n$ an integer, will evolve similarly, as the inflaton will have the same phase at the time the system becomes non-linear (remember $\chi(t) \propto e^{\mu t} \chi_{\rm i}$). In fact, if there were no inhomogeneous modes at all, the behaviour of the fields would be exactly the same for all $\chi_{\rm i}', \chi_{\rm i}$ related as in Eq.~(\ref{eq:ChiToChi}), as in this case only the phase information matters. As at the time of non-linearities the inhomogeneous modes are still small, we expect the field behaviour (and therefore the value of physical observables that depend on it) to repeat periodically in the space of initial values $\chi_{\rm i}$. This was indeed observed for curvature perturbations in \cite{Bond}. We have observed the same effect, but in the GW amplitude. We found that regions of high GW amplitude repeat log-periodically, as shown in Fig.~\ref{chiGWtot}.

\begin{figure}[t]
\begin{center}
\includegraphics[width=8.0cm]{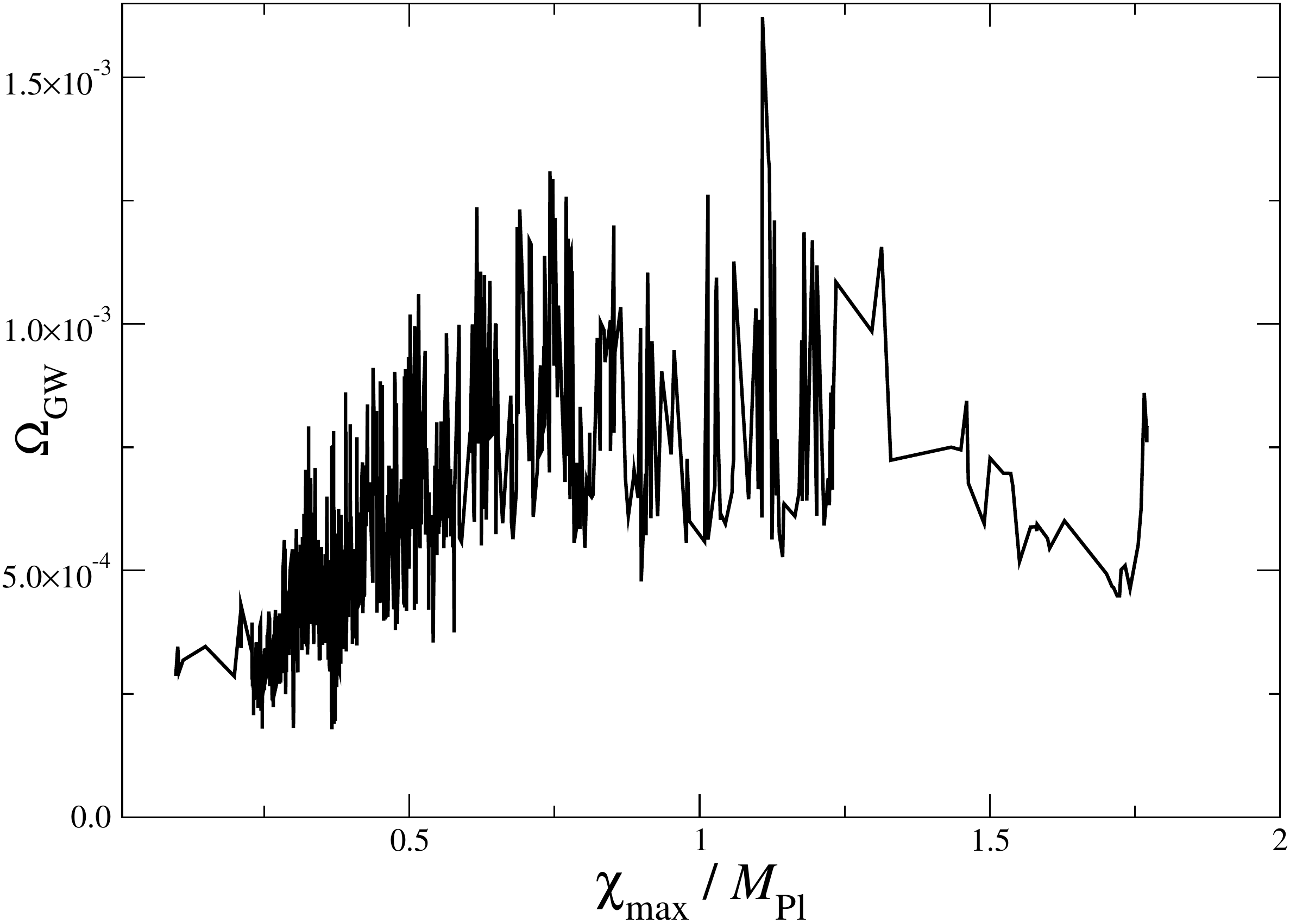}
\end{center}
\caption{The correlation between the maximum amplitude of the homogeneous part of $\chi$, $\chi_{\rm max}$, and the total GW energy in the simulation.}
\label{chimaxGW}
\end{figure}

To quantify how the GW production and the dynamics of $\chi$ are related, we studied how the maximum value the homogeneous field $\chi$ reaches during its evolution, $\chi_{\rm max}$, correlates with the amplitude of the final GW background. This maximum value indeed varies considerably between different $\chi_{\rm i}$, so it is clearly a good indicator that something is going on differently in the dynamics of $\chi$. Obviously the GWs are not sourced by the homogeneous field itself, but rather by its inhomogeneous modes. However, the latter are directly linked to the zero mode due to the transfer of energy between them during the non-linear stage. In Fig.~\ref{chimaxGW} we plot $\chi_{\rm max}$ against the total amount of GW energy, for the same simulations as in Fig.~\ref{chiGWtot}.

For small  $\chi_{\rm max}<1 M_{\rm Pl}$, we can see a clear correlation between the field dynamics and GW production: the more energy is deposited into the $\chi$ field, the more GW are being produced. This agrees with our findings from Fig.~\ref{GWsources}, showing that $\chi$ is responsible for the shape and amplitude of the GW spectra. For high $\chi_{\rm max}\gtrsim 1.2M_{\rm Pl}$, the correlation seems to turn around, and less GW are being produced, although due to the lack of data in this high $\chi_{\rm max}$ region, it is difficult to make a proper quantitative statement. Using a smaller lattice, $\tilde{L}=25$, we were able to find values of $\chi_{\rm i}$ which led to a very high field value $\chi_{\rm max}\gtrsim 5M_{\rm Pl}$, and for these the GW amplitude was highly suppressed.
A potential reason for the suppression might be that for low enough $\chi_{\rm max}$, the homogeneous $\chi(t)$ field oscillates fast enough to transfer energy to the inhomogeneous modes during the time of GW production, thus sourcing more GWs when more energy can be deposited. For very large $\chi_{\rm max}$, however, $\chi(t)$ only does very few oscillations, and most of the energy is stored in the homogeneous mode, thus reducing the field gradients and correspondingly the GW production. 

As previously mentioned, the scalar field $\chi$ is also responsible for an additional, highly non-Gaussian contribution to the curvature perturbation $\delta N$ \cite{rajantie, Bond}. In general, one can therefore expect a correlation between the GW anisotropies we studied and non-Gaussian features in the CMB. Because our numerical accuracy was not high enough to measure the curvature perturbations, we were unable to investigate this correlation.

\section{Conclusions}
\label{sec:conclusions}
We have shown that when preheating takes place via parametric resonance of a light field $\chi$, with mass $m_\chi \ll H_*$, the energy density in the expected background of GWs is generically anisotropic. During inflation $\chi$ acquires an almost scale-invariant spectrum, such that its amplitude on superhorizon scales varies randomly from one Hubble patch to another at the onset of preheating. The initial amplitude $\chi_{\rm i}$ at each causally disconnected patch sets up the initial conditions for preheating. In particular, it affects the late non-linear dynamics during which GWs are primarily produced, and therefore even small change in $\chi_{\rm i}$ can lead to a very different GW amplitude.
As a result, the GW background from preheating varies on superhorizon scales due to the variation of $\chi$. 

In this paper we have focused on the massless preheating model ${1\over4}\lambda\phi^4 + {1\over2}g^2\phi^2\chi^2$, as its conformal invariance makes it numerically very convenient since it allows us to re-scale away the expansion of the Universe. We have fixed the couplings to $g^2/\lambda = 2$, which guarantees that during the last e-folds of inflation $\chi$ is light, i.e.~its mass verifies $m_\chi = g\phi < H_*$. Additionally, this choice of couplings ensures that the very long wavelengths of $\chi$ are the fastest growing modes during the initial linear stages of parametric resonance.
When non-linearities become important, these long wavelengths modes transfer energy into the short ones, such that the dynamics of the latter are affected by the amplitude of the long-wavelength mode which is directly proportional to the initial value $\chi_{\rm i}$. Therefore, the non-linear stage, during which most power is added into the GWs, occurs differently for different causally disconnected regions.  This picture is supported by simulations with $g^2/\lambda = 1$ and $6$, in which the homogeneous mode was not amplified by the resonance and, consequently, no anisotropy was produced. Note that in other models (not massless) the zero mode is often amplified during part of the preheating process for a broad range of coupling values, so the effect is not expected to be limited to a very narrow range of the parameter space.

For the specific choice of couplings we considered, $g^2/\lambda = 2$, we found significant anisotropy, with dipolar and quadrupolar modulation representing variations of $\sim1\%$ in the GW energy density. The angular power spectrum is indeed flat, in the sense that $l(l+1)C_l$ is constant. 
This flat spectrum is a characteristic that can also be expected in other scenarios where anisotropies are present in the corresponding GW background, as long as the mechanism to produce them is similar to the one we have discussed in this paper. 
This includes other, more realistic theories of preheating via parametric resonance, and also other models with light scalars in which non-equilibrium processes generate gravitational waves, for example the resonant curvaton model~\cite{Enqvist:2008be,Chambers:2009ki,D'Onofrio:2012qy}. Of particular phenomenological significance will be models in which the Standard Model Higgs field plays the role of the light scalar, as for instance in~\cite{HiggsCurvaton}. \\\\

\section*{Acknowledgments}
This research was funded by STFC grants
ST/J000353/1 and ST/F007027/1, and the Royal
Society International Joint Project JP100273. DGF
is supported by the Swiss National Science Foundation. The simulations were carried out using the
COSMOS@DiRAC facility which is supported by
STFC/DBIS UK.

\end{document}